\pdfoutput=1
\documentclass[pre,onecolumn]{revtex4-2}
\usepackage{graphicx}
\usepackage{natbib}
\usepackage{amsmath,amssymb}
\usepackage{graphicx}
\usepackage{subcaption}

\newcommand{\be}{\begin{equation}}
\newcommand{\ee}{\end{equation}}

\begin{document}

\title{Spatial Wave Pattern in Locally Coupled Kuramoto Model}
\author{Yi Yu}
\affiliation{Center for Quantitative Biology, Peking University, Beijing 100871, China}

\begin{abstract}

  The Kuramoto model is a commonly used mathematical model for studying synchronized oscillations in biological systems, with its temporal synchronization properties well studied. However, the properties of spatial waves have received less attention. This paper investigates the spatial waves formed by locally coupled oscillators arranged in an $n\times n$ grid. Numerical simulations show that directional waves can form when the system exhibits heterogeneity, while spiral waves can arise in homogeneous systems. Interestingly, both wave patterns remain stable under minor noise disturbances. To explain the properties of the spatial wave pattern, starting from the simplest case of a $2\times 2$ grid, we analytically calculate the phase differences between oscillators to discuss the formation of wave patterns in the system. We then apply this method to compute the stable and saddle points and corresponding wave patterns of some $n\times n$ grid cases and discuss their stability. Furthermore, linear approximation reveals that the wave pattern under noise is the noiseless wave pattern plus its first-order approximation, indicating that the wave pattern remains stable within a certain range of noise. These results suggest that the necessary condition for directional wave propagation in biological systems is the presence of heterogeneity that far exceeds noise. In contrast, the disappearance of heterogeneity may induce spiral waves, often corresponding to disease states.

\end{abstract}

\maketitle

\section{Introduction}

Winfree's novel paper\cite{Winfree1967-mr} introduced the population oscillator model into biological research, mainly used to study the phenomenon of synchronized oscillations that frequently occur in biological systems. The Kuramoto model\cite{Strogatz2000-ci, Acebron2005-wz} is the most well-known population oscillator model used to study this phenomenon. In biological systems, synchronized oscillations are often accompanied by spatial wave propagation, such as the propagation of electrical signals in the nervous system\cite{Ermentrout2001-ar} or calcium waves in the heart\cite{Bers2008}. Some specific wave patterns, such as spiral waves, are corresponded to disease states.\cite{Ermentrout2001-ar}  Our lab recently focused on synchronized calcium oscillations in pancreatic islet cells.\cite{Ren2022-bo} The pancreas is an organ responsible for maintaining blood sugar homeostasis. After a meal, the rise in blood sugar levels stimulates the beta cells in the pancreatic islets to secrete insulin. During this process, the cytoplasm calcium concentration in beta cells also periodically increases synchrony, which is crucial for hormone release. Therefore, figuring out the underlying mechanism of synchronized calcium oscillations is essential. 

Our data show that calcium waves in healthy islets led by pacemaker cells propagate through whole islets. However, spiral waves can appear in the pancreatic islet model under high-fat diet-induced pre-diabetic conditions. Since calcium in single beta cells oscillates simultaneously under high glucose stimulation, beta cells in islets are electrically coupled with local neighbors by gap junction, it is proper to use a locally coupled Kuramoto model to study spatial wave patterns. (Details see Section II) Numerical simulations showed that wave propagation could be formed from this cluster when the system exhibits spatial heterogeneity, such as a cell cluster with a faster intrinsic frequency. When this heterogeneity disappears, spiral waves can emerge in the system. Moreover, these two types of spatial wave patterns are robust to noise. It is qualitatively similar to the observation in experiments, which encouraged us to discuss the pattern formation in this system theoretically. 

The temporal synchronization \cite{Strogatz2000-ci, Hoang2015-bh, Bick2020-mi} and spatial wave properties \cite{Paullet1994, Avalos2009-nc, Sieber2011-mr, Udeigwe2015, Jason2019, Sarkar2021-sl} are well-studied in Kuramoto models. Considerable discussion has been devoted to traveling wave propagation and stability in the Locally Coupled Kuramoto Model in one-dimensional settings, as reported in previous works\cite{Tilles2011-mp, Sieber2011-mr}, while the focus in two-dimensional systems has been on the complete phase-lock state (a trivial solution in which there is no phase difference in the system) and spiral wave patterns. Ermentrout and Paullet's seminal work \cite{Paullet1994} first reported a globally stable rotation solution to the locally coupled Kuramoto model by introducing proper symmetry, which has since been applied in subsequent works to regular graphs \cite{Udeigwe2015} and discussed extensively in infinite systems \cite{Jason2019}. Aside from discussing the rotation solution in terms of symmetry, another approach involves simplifying the Kuramoto model into other two-dimensional models, such as considering its continuum limit and approximating it as a reaction-diffusion model\cite{Avalos2009-nc}. Alternatively, it can be integrated into the XY model\cite{Sarkar2021-sl}, with the vortex structure (spiral wave pattern) observed in the Kuramoto model being explained through the Kosterlitz-Thouless (KT) phase transition.

While these impressive works have demonstrated various wave patterns in the Kuramoto model, this article is a complementary to the field that introduces a new method for calculating wave pattern formation in the locally-coupled Kuramoto model. The method begins with a $2\times 2$ grid, calculating analytical solutions for the phase differences between oscillators under arbitrary conditions. It also discusses how introducing bilateral and quadrilateral symmetry into the degenerate form can result in rotated equilibrium states, respectively. Additionally, the method is applied to higher-order grids, providing analytical solutions for directional wave propagation in a $3\times 3$ grid and spiral wave patterns in a $4\times 4$ grid. Notably, based on discussions regarding the degenerate form in the $2\times 2$ grid and detailed calculations using the new method, this work not only reproduces the stable solution with quadrilateral symmetry reported in \cite{Paullet1994} (referred to as \textbf{W1} below), but also derives unstable equilibrium solutions with bilateral symmetry. Moreover, they are identified as 1st-order saddle points by calculating the eigenvalues of the Jacobian matrix. On the other hand, this article attempted to estimate the influence of biological noise. The linear approximation method gives a good approximation of the impact of biological noise on wave patterns, highlighting the significance of heterogeneity in biological systems, which far exceeds that of noise.

\section{Numerical Simulation}

\begin{figure}[htbp]
    \centering
    \includegraphics[width=1\textwidth]{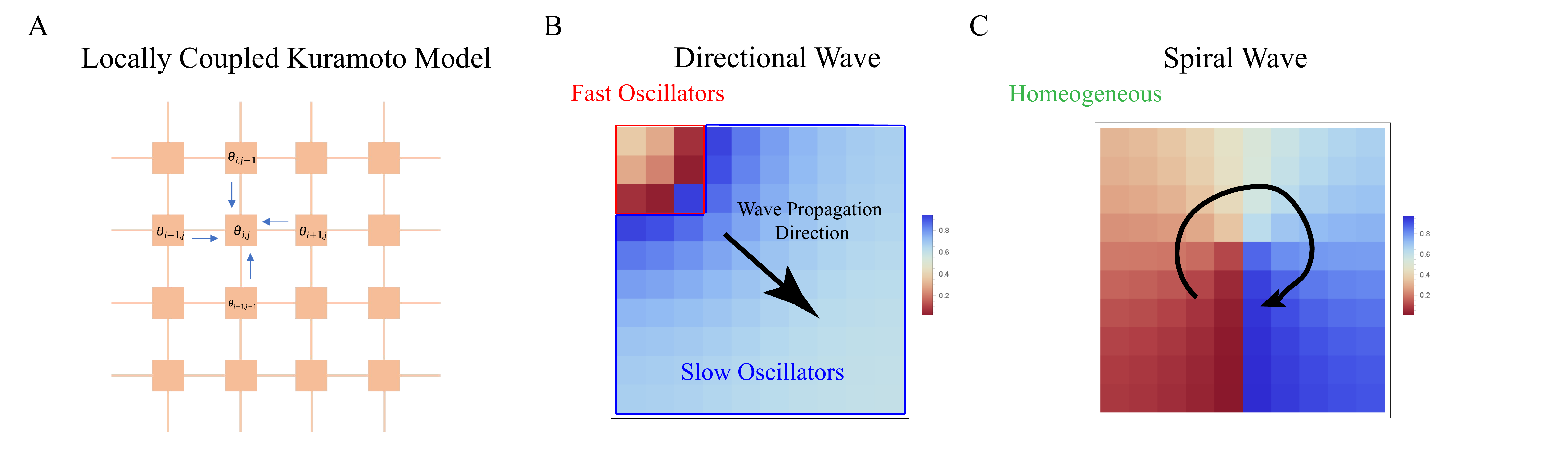}
    \caption{ Typical wave patterns in locally coupled Kuramoto model. }
    \label{fig:central illustration}
\end{figure}

The locally-coupled model discussed in this article refers to oscillators arranged on an $n \times n$ grid, where each oscillator only connects to its neighboring oscillators, (FIG. \ref{fig:central illustration} A). Since the number of cells in the biological system is finite and has boundaries, the model considers finite oscillators and does not have periodic boundaries, i.e., oscillators at the corner only have two neighbors. Our model assumes two types of oscillators in the system with different intrinsic frequencies to simulate the calcium waves observed in the pancreatic islets. The faster oscillators, which arrange at corners, act as pacemaker cells to trigger directional wave propagation (FIG. \ref{fig:central illustration} B). However, replacing these fast oscillators with slow ones, the system ocassionally exhibits spiral wave patterns. (FIG. \ref{fig:central illustration} C)

\begin{figure}[htbp]
    \centering
    \includegraphics[width=\textwidth]{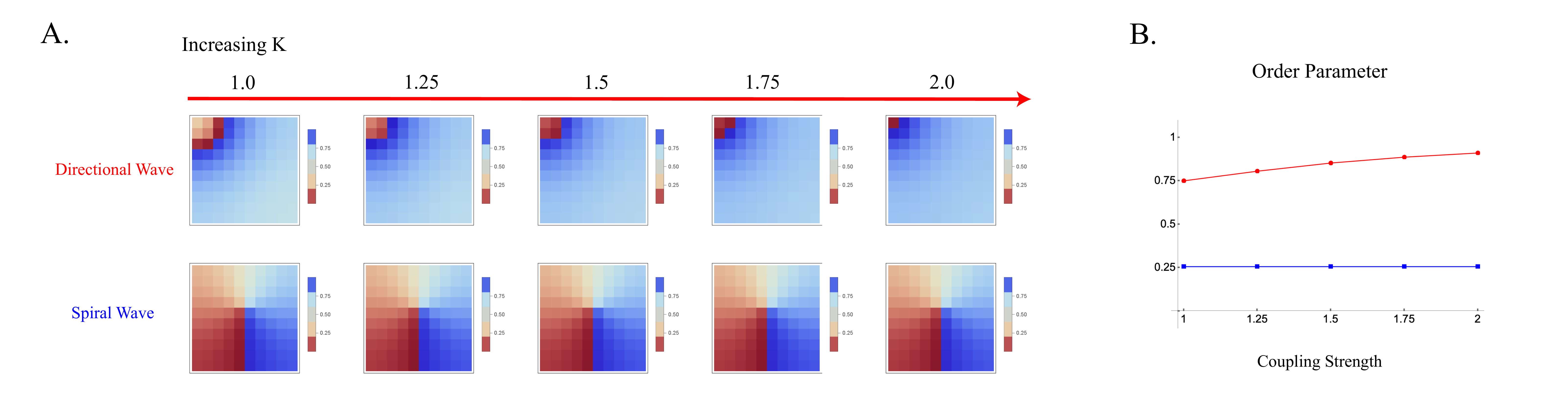}
    \caption{ Wave pattern changes with increasing coupling strength $K$}
    \label{fig:increasing k}
\end{figure}

Calculation of order parameters of the system reveals that the directional wave approaches a value of 1, while the spiral wave corresponds to a small value. (FIG. \ref{fig:increasing k}B, the initial point of each line) Moreover, the sensitivity of these two wave patterns to changes in parameters is markedly different. In the Kuramoto model, the coupling constant k between oscillators is the most crucial parameter. Intuitively, as the strength of the coupling increases, the wave propagation speed should increase, and the phase difference between oscillators in the system should decrease. Numerical simulations confirm that the directional wave behaves as expected: as the coupling strength increases, the maximum phase difference between oscillators in the system decreases and the wave speed increases. (FIG. \ref{fig:increasing k}A, directional wave) However, the behavior of the spiral wave exhibits little change under these conditions. (FIG. \ref{fig:increasing k}A, spiral wave) This can also be observed from the changes in the order parameter: the order parameter of the directional wave increases as the coupling strength increases (FIG. \ref{fig:increasing k}B, red line), whereas the order parameter of the spiral wave remains relatively unchanged(FIG. \ref{fig:increasing k}B, blue line).

\begin{figure}[htbp]
    \centering
    \includegraphics[width=0.8\textwidth]{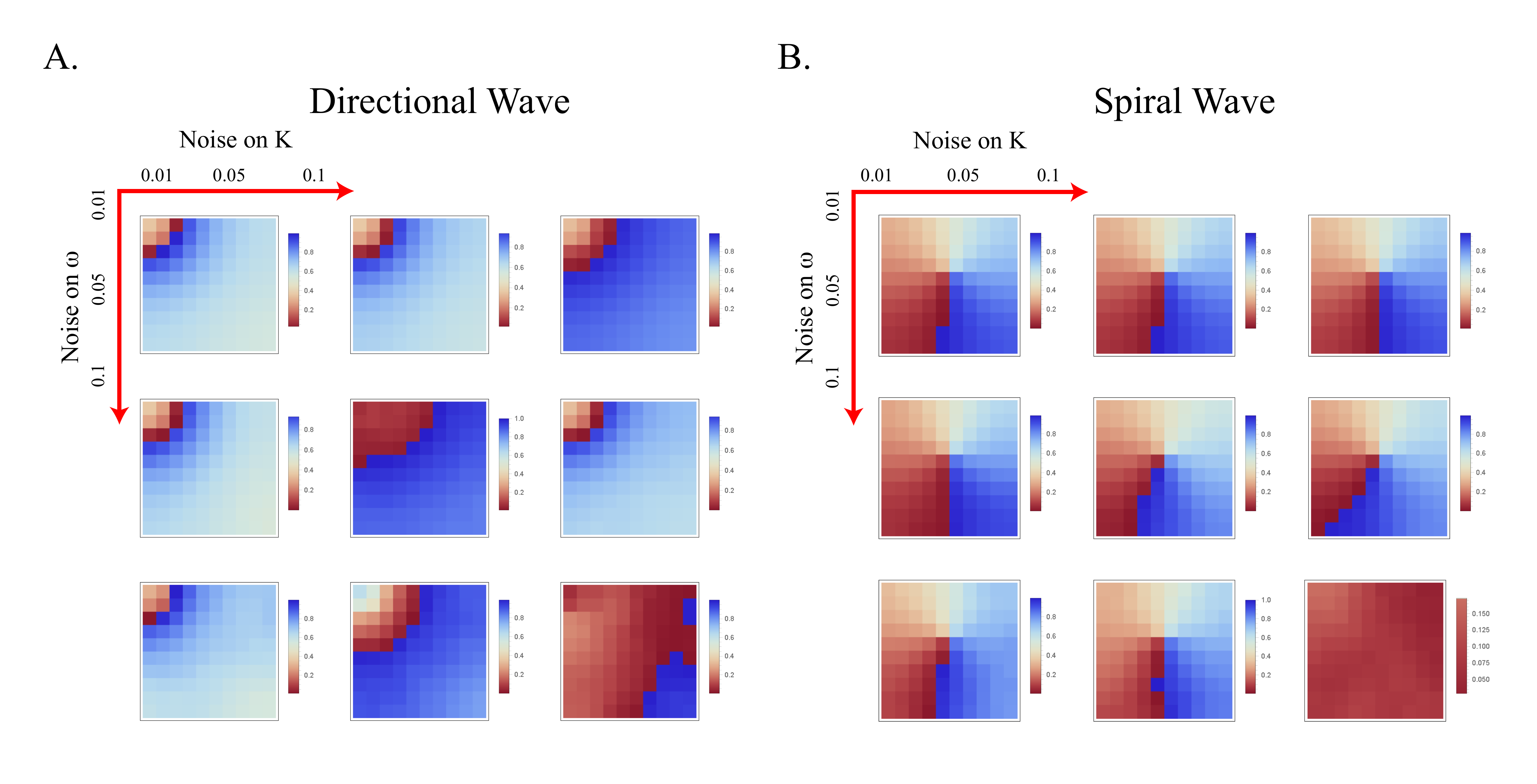}
    \caption{ Wave pattern changes with increasing noise on $K$ and $\omega$}
    \label{fig:increasing noise}
\end{figure}

In actual biological systems, noise cannot be ignored, and parameters in the system often exhibit some degree of randomness. Therefore, simulations with noise were performed in this study by adding Gaussian noise to the intrinsic frequency $\omega$ and coupling strength $k$, where the intensity of the noise is determined by the variance $\sigma^2$. The simulation results indicate that adding small levels of noise to either the intrinsic frequency or coupling strength does not significantly alter the wave patterns of the directional or spiral waves (as shown in FIG. \ref{fig:increasing noise}A,B, the first and second rows and columns). However, when both are subjected to relatively high levels of noise, the original wave patterns may be disrupted (as shown in FIG. \ref{fig:increasing noise}A,B, the third row and third column). This suggests that both types of wave patterns are stably present and possess limited ability to resist noise.

\section{Condition of Equilibrium State}

Consider a symmetrically-coupled Kuramoto model (eq.\ref{eq: kuramoto model}),

\be 
 \frac{d\theta_i}{dt} = \omega_i + \sum_{j\in \Lambda_i} k_{ji}\sin(\theta_j-\theta_i) \label{eq: kuramoto model} 
\ee

with average rotation speed \(\overline{\omega}\). $\Lambda_i$ contains the indices of neighbor oscillator $j$ that are adjacent to oscillator $i$.

In the context of the oscillator model, it is common for the natural frequency of oscillators to be greater than zero, resulting in the derivative of $\theta_i$ with respect to time ($\frac{d\theta_i}{dt}$) being non-zero at the equilibrium state. However, even though the system is not in a stable state in the traditional sense, the oscillators may still reach a state of phase locking, where the relative motion between oscillators is zero. In this study, the equilibrium state refers to the phase-locked state, where all oscillators are synchronized such that for any two oscillators $\theta_i$ and $\theta_j$, $\lim_{t \to \infty} \theta_j-\theta_i = C_{ji}, \forall i,j$ holds, where $C_{ji}$ is a constant. The following lemma provides evidence that the phase-locked equilibrium state of oscillators can be effectively represented by a hypothetical "average oscillator" phase difference augmented by a constant term. Proof of \textbf{Lemma 1} is stated in Appendix.A. 

\textbf{Lemma 1} Dynamic system (eq.1) reaches an equilibrium state
equivalent to that all oscillators \(\theta_i\) are phase-locked to
average rotation \(\overline{\omega}t\). In other words,
\(\theta_i -\overline{\omega}t \equiv \Omega_i, \forall i\).

\section{Typical Wave Pattern in The $2\times 2$ Grid} 

The $2\times 2$ square grid is the smallest unit of a square grid, with each oscillator at the corner interacting only with its two neighboring oscillators. In this section, I will first present the analytical solution for the phase shift between oscillators when the system reaches equilibrium in a 2x2 grid. Then, I will demonstrate how this method can be used to analyze a general nxn grid. Finally, I will illustrate the possible complex dynamics of this simple dynamical system.

\subsection{Analytic solution to the $2\times 2$ grid}

\begin{figure}[htbp]
    \centering
    \includegraphics[width=0.8\textwidth]{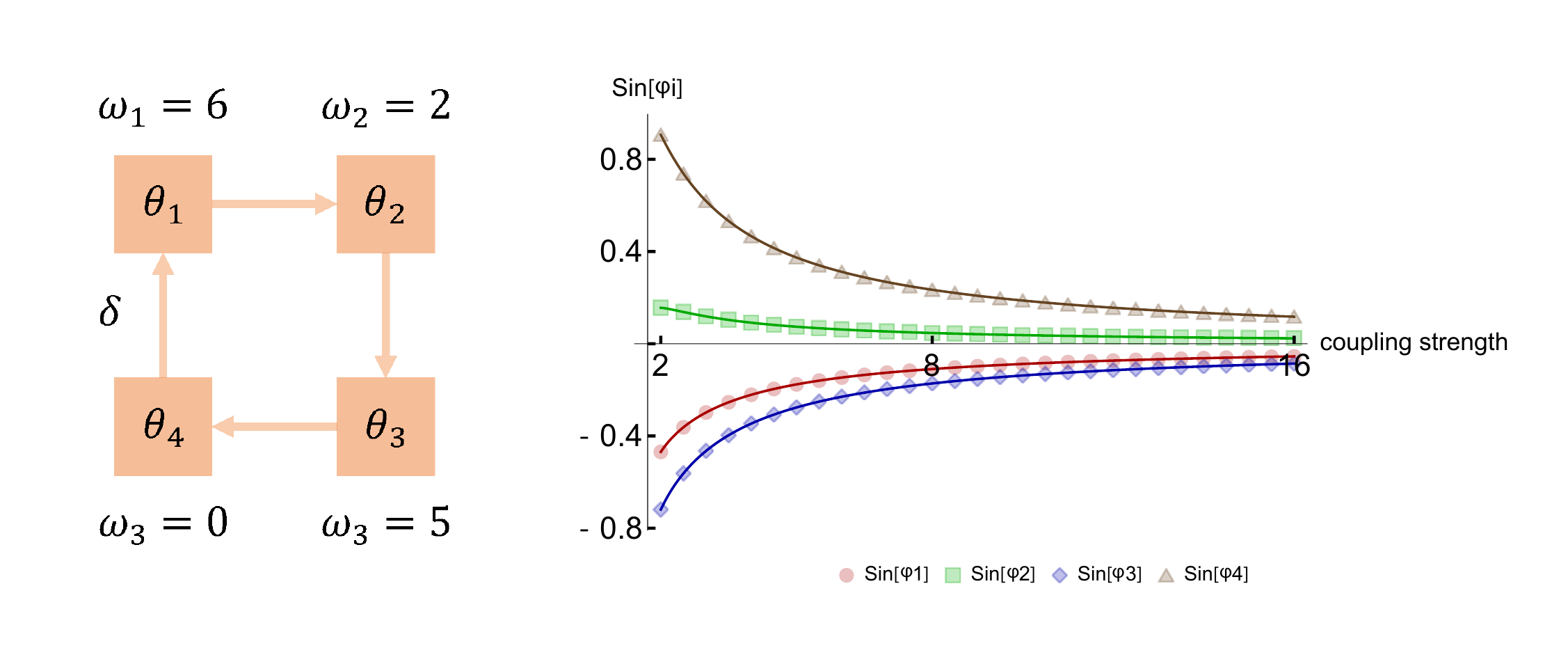}
    \caption{$\sin(\phi_i)$ decreases with increasing $k$. Dots: numerical simulation. Curves: analytic solution}
    \label{fig:resultIII11}
  \end{figure}

The explicit equation for this system is shown below, in which $\phi_1 = \theta_2-\theta_1,\phi_2 = \theta_3-\theta_2,\phi_3 = \theta_4-\theta_3,\phi_4 = \theta_1-\theta_4$, as it is shown in FIG. \ref{fig:resultIII11}A.

\begin{equation} \label{eq:def}
 \begin{aligned}
 \frac{d\theta_1}{dt} = \omega_1+k(\sin(\phi_1)-\sin(\phi_4))\\
 \frac{d\theta_2}{dt} = \omega_2+k(\sin(\phi_2)-\sin(\phi_1))\\
 \frac{d\theta_3}{dt} = \omega_3+k(\sin(\phi_3)-\sin(\phi_2))\\
 \frac{d\theta_4}{dt} = \omega_4+k(\sin(\phi_4)-\sin(\phi_3))
 \end{aligned}
\end{equation}

By subtracting these equations pairwise, we obtain the dynamics equation that is solely related to the phase difference, as demonstrated below

\begin{equation}
 \begin{aligned}
 \frac{d\phi_1}{dt} = \omega_2-\omega_1+k(\sin(\phi_2)+\sin(\phi_4)-2\sin(\phi_1))\\
 \frac{d\phi_2}{dt} = \omega_3-\omega_2+k(\sin(\phi_3)+\sin(\phi_1)-2\sin(\phi_2))\\
 \frac{d\phi_3}{dt} = \omega_4-\omega_3+k(\sin(\phi_4)+\sin(\phi_2)-2\sin(\phi_3))\\
 \frac{d\phi_4}{dt} = \omega_1-\omega_4+k(\sin(\phi_1)+\sin(\phi_3)-2\sin(\phi_4))
 \end{aligned}
\end{equation}

This dynamical system's equilibrium state comprises the following equations, which were written in a matrix form.
\be
 \begin{bmatrix}-2& 1& 0&1\\ 1 & -2 & 1 & 0 \\ 0 & 1 & -2 & 1 \\ 1 & 0 & 1 & -2 \end{bmatrix} \begin{bmatrix}\sin(\phi_1) \\ \sin(\phi_2) \\ \sin(\phi_3) \\ \sin(\phi_4) \end{bmatrix} = \begin{bmatrix} \frac{\omega_1-\omega_2}{k} \\  \frac{\omega_2-\omega_3}{k} \\ \frac{\omega_3-\omega_4}{k} \\ \frac{\omega_4-\omega_1}{k} \end{bmatrix}
\ee

This is the solution to the equation, and because the coefficient matrix is rank deficient, the answer to the equation possesses one degree of freedom $\delta$.
\be
 \begin{bmatrix}\sin(\phi_1) \\ \sin(\phi_2) \\ \sin(\phi_3) \\ \sin(\phi_4) \end{bmatrix} = \begin{bmatrix}\frac{-3\omega_1+\omega_2+\omega_3+\omega_4}{4k} \\ \frac{-\omega_1-\omega_2+\omega_3+\omega_4}{2k} \\ \frac{-\omega_1-\omega_2-\omega_3+3\omega_4}{4k} \\ 0 \end{bmatrix} +\delta
\ee

To determine the specific value of this degree of freedom, it is necessary to consider a natural constraint on $\phi_i$:
\be\label{phisum}
 \phi_1+\phi_2 +\phi_3+\phi_4 = 0
\ee

From this equatithree new equations can be obtained by selecting two variables and moving them to the other sideline, taking the sine on both sides.
\begin{equation}
 \begin{aligned}
 \sin(\phi_1+\phi_2) = -\sin(\phi_3+\phi_4)\\
 \sin(\phi_1+\phi_3) = -\sin(\phi_2+\phi_4)\\
 \sin(\phi_1+\phi_4) = -\sin(\phi_2+\phi_3)
 \end{aligned}
\end{equation}

By expanding the terms in these equations and rearranging them into matrix form, it is possible to obtain the following:
\be
 \begin{bmatrix}\cos(\phi_2)& \cos(\phi_1)& \cos(\phi_4)& \cos(\phi_3)\\ \cos(\phi_3) & \cos(\phi_4) & \cos(\phi_1) & \cos(\phi_2) \\ \cos(\phi_4) & \cos(\phi_3) & \cos(\phi_2) & \cos(\phi_1) \end{bmatrix} \begin{bmatrix}\sin(\phi_1) \\ \sin(\phi_2) \\ \sin(\phi_3) \\ \sin(\phi_4) \end{bmatrix} = \begin{bmatrix}0 \\0 \\0 \end{bmatrix}
\ee

The intuitive interpretation of this linear equation is that the vectors comprised of the cosine terms as coefficients are situated within the null space of the vector constituted by the sine terms.
\be
 \begin{bmatrix}\cos(\phi_2)\\ \cos(\phi_1)\\ \cos(\phi_4)\\ \cos(\phi_3)\\  \end{bmatrix} ,
 \begin{bmatrix}\cos(\phi_3)\\ \cos(\phi_4)\\ \cos(\phi_1)\\ \cos(\phi_2)\\  \end{bmatrix}, 
 \begin{bmatrix}\cos(\phi_4)\\ \cos(\phi_3)\\ \cos(\phi_2)\\ \cos(\phi_1)\\  \end{bmatrix} \in 
 \text{nullspace}[\begin{bmatrix}\sin(\phi_1) \\ \sin(\phi_2) \\ \sin(\phi_3) \\ \sin(\phi_4) \end{bmatrix} ]
\ee

By deriving the expression for the sin component vector at the outset, its corresponding null space can also be represented using a free variable $\delta$.

\be
 \text{nullspace}[\begin{bmatrix}\sin(\phi_1) \\ \sin(\phi_2) \\ \sin(\phi_3) \\ \sin(\phi_4) \end{bmatrix} ] =\{ 
    \begin{bmatrix}\eta_1\\ 1 \\ 0 \\ 0 \end{bmatrix} ,
    \begin{bmatrix}\eta_2\\ 0 \\ 1 \\ 0 \end{bmatrix} ,\begin{bmatrix}\eta_3 \\ 0 \\ 0 \\ 1 \end{bmatrix}  \}
\ee

In which,

\begin{equation}
 \begin{aligned}
  \eta_1 &= \frac{2(\omega_1+\omega_2-\omega_3-\omega_4-2k\delta)}{-3\omega_1+\omega_2+\omega_3+4k\delta} \\
  \eta_2 &=\frac{\omega_1+\omega_2+\omega_3-3\omega_4-4k\delta}{-3\omega_1+\omega_2+\omega_3+4k\delta} \\
  \eta_3 &= \frac{-4k\delta}{-3\omega_1+\omega_2+\omega_3 +\omega_4+4k\delta}
 \end{aligned}
\end{equation}

As a four-dimensional vector with a rank of 1, the null space has a dimensionality of three, aligning with the three vectors formed bycombiningf cosine terms. Here is an example:

\be
 \begin{bmatrix}\cos(\phi_2)\\ \cos(\phi_1)\\ \cos(\phi_4)\\ \cos(\phi_3)\\  \end{bmatrix}= \cos(\phi_1) \begin{bmatrix}\eta_1\\ 1 \\ 0 \\ 0 \end{bmatrix} +  \cos(\phi4) \begin{bmatrix}\eta_2\\ 0 \\ 1 \\ 0 \end{bmatrix} +  \cos(\phi_3) \begin{bmatrix}\eta_3\\ 0 \\ 0 \\ 1 \end{bmatrix}
\ee

Therefore it gives that
\be
 \cos(\phi_2) = \cos(\phi_1)\eta_1 + \cos(\phi_4)\eta_2 + \cos(\phi_3)\eta_3
\ee

Expanding the cosine vectors in the nullspace can obtain the equation relationship between $ \cos(\theta_i) $ can be obtained. Rearranging this equation relationship yields the following matrix form.

\be
 \begin{bmatrix}
 \eta_1& -1 & \eta_3 & \eta_2 \\ 
 \eta_2 & \eta_3 & -1 & \eta_1 \\
 \eta_3 & \eta_2 & \eta_1 & -1
 \end{bmatrix}
 \begin{bmatrix}\cos(\phi_1)\\ \cos(\phi_2)\\ \cos(\phi_3)\\ \cos(\phi_4)\\  \end{bmatrix}
 = \begin{bmatrix}0\\ 0\\ 0\\ 0\\  \end{bmatrix}
\ee

The linear equation obtained provides the conditions satisfied by $\cos(\theta_i)$, namely that it lies within the null space of the coefficient matrix constructed from $\eta_i$. Given that $\eta_i$ is represented by the free variable $\delta$ and that $\cos(\theta_i)$ can also be expressed as a function of $\delta$ through the essential formula $\cos(\theta)^2=1-\sin(\theta)^2$, a polynomial equation solely dependent on $\delta$ can be derived. Notably, this equation is a polynomial of degree at most 4, indicating the existence of analytical solutions. The explicit form is complex to show, coordinates of the nullspace vector is denoted as $c_i$ and the exact values are given in Appendix.B. That is, for example:

\begin{align}\label{eq:cos}
 \frac{\cos(\theta_1)}{\cos(\theta_4)} = \frac{c_1}{c_4}
\end{align}

Finally, $\cos(\phi_i)^2 = 1-\sin(\phi_i)^2$ takes back into eq.\ref{eq:cos}, and have:
\be
 \frac{1-\sin(\phi_1)^2}{1-\delta^2} = \frac{c_1^2}{c_4^2}
\ee

This fractional equation, after being simplified, becomes a quartic equation in terms of $\delta$.

Although the general solution form may be highly complex, this method provides a convenient approach to obtaining solutions for specific cases. Here are some examples where for the sake of simplicity, the coupling coefficient $k$ is assumed to be a constant $k$. For simplicity, $\omega_4$ is assumed to be zero. 

~\ 

1. $\omega_1 = \nu, \omega_2 = \rho + \nu, \omega_3 = \rho $.
   \be
    \frac{16(4 \mathrm{k} \delta+\rho-v)(-2 \mathrm{k} \delta+v)^2(-\rho+v)^3(\rho+v)^2}{k^2} = 0
   \ee

2. $\omega_1 = \rho, \omega_2 = \lambda\rho, \omega_3 = \rho $.
   \be
    -\frac{\lambda \rho^2(\lambda \rho-2 \rho)^3(8 k \delta+\lambda \rho-2 \rho)(-8 k \delta+\lambda \rho+2 \rho)^2}{16 k^2} = 0
   \ee

3. $\omega_1 = \kappa, \omega_2 = 2\kappa, \omega_3 = 3\kappa $.
   \be
    -256 \rho^4\left(16 k^2 \delta^2\left(-1+\delta^2\right)+16 k \delta^3 \rho+\left(1+3 \delta^2\right) \rho^2\right) =0
   \ee

4. $\omega_1 = \nu, \omega_2 = 0, \omega_3 = \rho$.
   \be
    -\frac{(\rho-3 v)(8 \mathrm{k} \delta+\rho-3 v)^3\left(\rho^2-v^2\right)^2}{16k^2} = 0
   \ee
~\

To verify the validity of the solution, I present a numerical calculation case. In this example, we assume $\omega_1=6, \omega_2=2, \omega_3=4, \omega_4=0$. The dots in the graph represent the $\sin(\phi_i)$ values at equilibrium obtained from the numerical calculation, while the curve is the analytical solution obtained from the formula. It can be seen that the two perfectly match each other, with the coupling coefficient step size of 0.5 for each data point.

\subsection{Extended $2\times 2$ grid}

This method gives a new way to calculate analytic phase shifts between neighbor oscillators. Consider an $n \times n$ grid with random connections and random intrinsic frequencies. By \textbf{lemma 1}, the equilibrium state of this system is determined by:

\be
\omega_i - \overline{\omega} + \sum_{j\in\Lambda} K_{ji}\sin(\phi_{j,i})=0
\ee

Where $\phi_{j,i} = \phi_j-\phi_i$. It contains $n^2$ equations while the number of variables is $2n(n-1)$. Moreover, the summation of all equations is $0$, which indicates one free variable. These equations have at least $(n-1)^2$ free variables. On the other hand, the $n \times n$ grid generates $(n-1)^2$ two-by-two grids, which consist o f $\phi_{i,j},i,j = 1,2,\dots,n$. Replacing $\phi_{i,j}$ into two-by-two grids will have $(n-1)^2$ equations for $(n-1)^2$ variables—theoretically, all of it could be solved analytically. 

\textbf{Lemma 2} if the equilibrium states of four oscillators following the two-by-two grids are $\sin(\phi_i) = \delta_i,i = 1,2,3,4$, then cosine of the phase shift vector $\begin{bmatrix}\cos(\phi_1)\\ \cos(\phi_2)\\ \cos(\phi_3)\\ \cos(\phi_4)\\  \end{bmatrix}$ is parallel to:

\begin{align}
 \begin{bmatrix}
 -\delta_1^3-2\delta_2\delta_3\delta_4+\delta_1(\delta_2^2+\delta_3^2+\delta_4^2) \\
 \delta_1^2\delta_2-2\delta_1\delta_3\delta_4+\delta_2(-\delta_2^2+\delta_3^2+\delta_4^2)\\
 \delta_1^2\delta_3-2\delta_1\delta_2\delta_4+\delta_3(\delta_2^2-\delta_3^2+\delta_4^2)\\
 \delta_1^2\delta_4-2\delta_1\delta_2\delta_3+\delta_4(\delta_2^2+\delta_3^2-\delta_4^2)
 \end{bmatrix}
\end{align}

\subsection{Complex behaviors in a degenerate $2\times 2$ grid}

Although \textbf{Lemma 2} is a powerful tool to get analytic solution, it was not helpful to degenerate form, i.e., more than $(n-1)^2$ freedoms in some cases, and then the phase shift vector is parallel to a hyperplane instead of a specific direction, which actually helps the spiral wave formation. The equivalent condition of degenerate form for a $2\times 2$ grid is:

\be
 \det(\begin{bmatrix}
  \delta_1 & \delta _2\\
  \delta_4 & \delta_3 
 \end{bmatrix}) = 0
\ee

It is an inspiring example, considering a two-by-two grid with constraints that $\omega_i = 0$ and $\sin(\phi_1) = \sin(\phi_2) = \sin(\phi_3) = \sin(\phi_4)= \gamma$. It is quite a simple case but exhibits abundant dynamic behaviors. 

First, $\sin(\phi_i)=\gamma = 0$ is a trivial solution to this problem. Besides, the direction vector of $\begin{bmatrix}\cos(\phi_1)\\ \cos(\phi_2)\\ \cos(\phi_3)\\ \cos(\phi_4)\\  \end{bmatrix} $ is $ \begin{bmatrix} -1 \\ 1 \\0 \\0  \end{bmatrix}, \begin{bmatrix} -1 \\ 0 \\1 \\0  \end{bmatrix}, \begin{bmatrix} -1 \\ 0 \\0 \\1  \end{bmatrix}$ which implies various solutions other than the trivial one. By adding $2\pi$ to both sides of eq.\ref{phisum} and partitioning $2\pi$ equally among each $\phi_i$, the equation can be transformed into:

\begin{align}
 (\phi_1 + \frac{\pi}{2}) + (\phi_2 + \frac{\pi}{2}) + (\phi_3 + \frac{\pi}{2}) + (\phi_4 + \frac{\pi}{2}) = 2\pi
\end{align}

Let $\phi'_i = \phi_i + \frac{\pi}{2}$, where $i = 1, 2, 3, 4$. Then it follows that:

\be
\cos(\phi'_1) = \cos(\phi'_2) = \cos(\phi'_3) = \cos(\phi'_4)
\ee

The complex numbers $\xi_i$ are defined as:

\be
\xi_i = e^{i\phi'_i}
\ee

It is stated that all $\xi_i$ are identical or conjugates of each other, and:

\be
\xi_1\xi_2\xi_3\xi_4 = 1
\ee

Therefore, there are three possible scenarios:

\vspace*{5pt}

\textbf{S0. (Complete synchronization)} If $\xi_i$ are identical, then $\xi_i^4 = 1$. This results in $\phi_i' = \frac{\pi}{2}$, which corresponds to the trivial solution $\phi_i=0$. 

\vspace*{5pt}

\textbf{S1. (Quadrilateral rotation symmetry)} If one pair of conjugate $\xi_j$,  assuming that is $\phi_1$ and $\phi_2$, and the other two are identical, then $\xi_i^2\xi_j\overline{\xi_j}=\xi_i^2=1$. It follows that $\phi'_1 = \phi'_2 = \phi'_3  = 0 $ and $ -\phi'_4 $ equals $2\pi$ (\textbf{Clockwise}), or $\phi'_1 = \phi'_2= \pi$ and $\phi'_3  = -\phi'_4 $ equals $\pi$ or $-\pi$ (\textbf{Anti-clockwise}), since $\cos(\theta'_2)=\cos(\theta'_3)$.

\vspace*{5pt}

\textbf{S2. (Bilateral-like rotation symmetry)}  If there are two pairs of conjugate $\xi_i$, then $\xi_1\xi_2\xi_3\xi_4 = \xi_i\overline\xi_i\xi_j\overline\xi_j=1 $ is always true. Assume $\phi_1 = \theta^*$ then possible solutions overfull $\phi_2, \phi_3, \phi_4$ can be taken from any permutation of $\theta^*, -\theta^*, 2\pi - \theta^*$.

\vspace*{5pt}

\begin{figure}[htbp]
    \centering
    \includegraphics[width=0.75\textwidth]{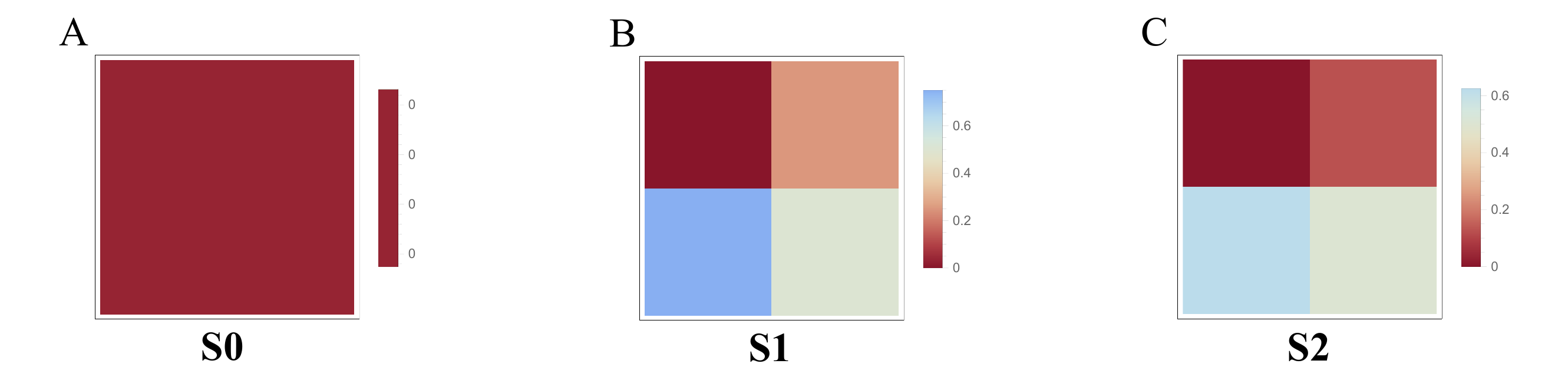}
    \caption{Wave pattern of a 2 by 2 grid. Left: \textbf{S0}.  Middle: \textbf{S1}.  Right: \textbf{S2}.}
    \label{fig:resultIII12}
  \end{figure}

The physical pictures in each scenario are clear: (1) The trivial solution states that all oscillators are completely synchronized without phase shifts. (2) The second case generates a rotation with phase shift $\frac{\pi}{2}$. (3) The last case generates a bilateral-like rotation symmetry. In particular, only the first case is a stable equilibrium state, while the other two are unstable. Replacing $\phi_4$ by $\phi_4 = -\phi_1-\phi_2-\phi_3$, this dynamic system is reduced, and the corresponding Jacobian becomes

\vspace*{0.5pt}

\footnotesize \begin{align}
 \begin{bmatrix}
 -2\cos(\phi_1)-\cos(\phi_4) & \cos(\phi_2)-\cos(\phi_4) & -\cos(\phi_4) \\ 
 \cos(\phi_1) & -2\cos(\phi_2) & \cos(\phi_3) \\   
 -\cos(\phi_4) & \cos(\phi_2)-\cos(\phi_4) & -2\cos(\phi_3)-\cos(\phi_4) 
 \end{bmatrix}
\end{align}

\normalsize For the first case, the eigenvalues of Jacobian are $-4,-2,-2$, which means the trivial solution is stable. For the case (b), the Jacobian becomes an all-zero matrix with all zero eigenvalues. For the case (c), there are three possible results. If diagonal phase shifts are conjugate, i.e., $\xi_1 = \bar{\xi_3}$ and $\xi_2 = \bar{\xi_4}$ , then the eigenvalues are $-2\sqrt{2}\lambda,2\sqrt{2}\lambda,0$. If neighbor phase shifts are conjugate, i.e., $\xi_1 = \bar{\xi_2}$ and $\xi_3 = \bar{\xi_4}$, then the eigenvalues are $-2\lambda,2\lambda,0$. All of them indicate a saddle equilibrium state.  

In the next section, I will show that all those dynamic behaviors emerge in a larger grid.

\section{Spatial Wave Pattern in $n\times n$ Grid} 

In this section, I will use \textbf{Lemma 1} and \textbf{Lemma 2} to provide explanations for some of the phenomena observed in numerical simulations.

\subsection{Directional wave}

To simplify the problem, consider a model in which only the oscillator located at the top left corner of an $n \times n$ grid has a larger rotational speed $\omega_h$, while all other oscillators have a lower speed $\omega_l$, and their difference is $\omega_h - \omega_l = \omega$. This means that the wave propagation is driven by this corner oscillator. 

\subsubsection{Analytic solution to the $3\times 3$ grid}

\begin{figure}[htbp]
    \centering
    \includegraphics[width=0.8\textwidth]{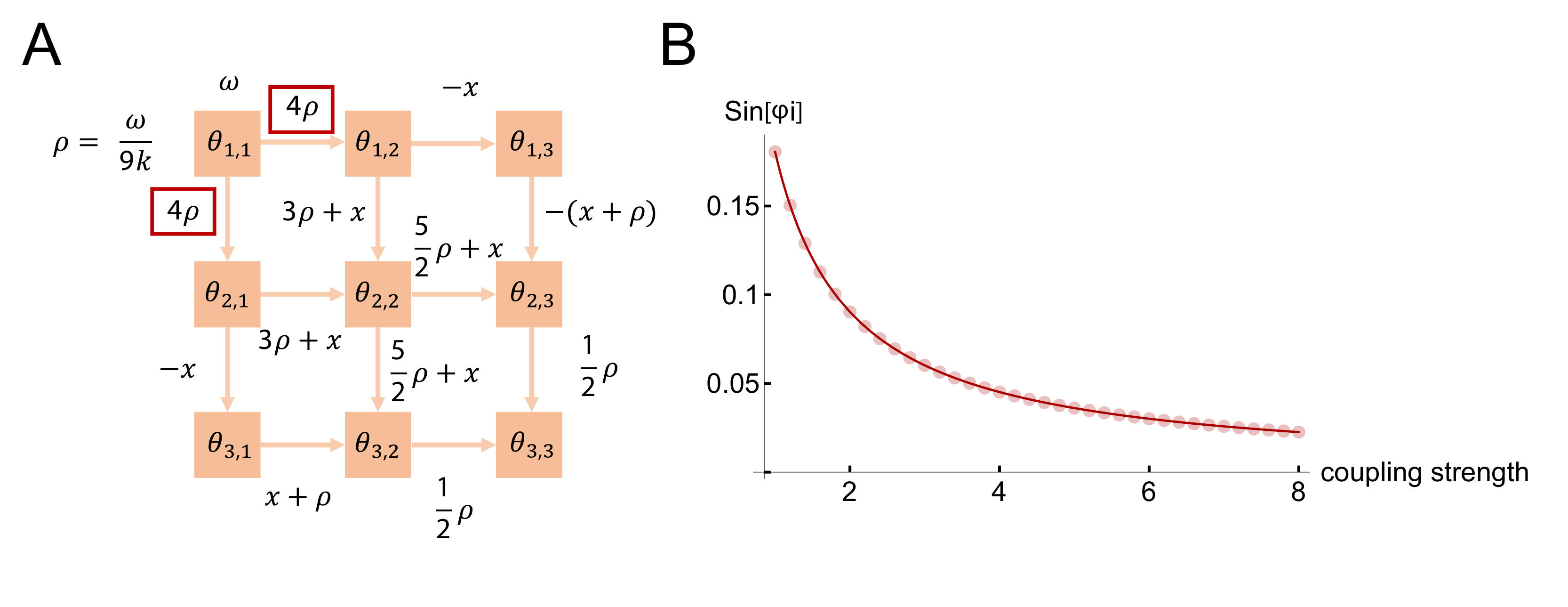}
    \caption{Phase shift $x$ decreases with increasing $k$. Red dots: numerical simulation. Red curve: analytic solution}
    \label{fig:resultIII21}
  \end{figure}

First of all, consider a simple example of a $3\times3$ grid of oscillators arranged as shown in FIG.\ref{fig:resultIII21}A. Using \textbf{Lemma 1}, the equilibrium condition for this system can be obtained as shown in FIG.\ref{fig:resultIII21}A. Additionally, assume that the coupling strengths, $k_{ij}=k$, are identical. Now, applying \textbf{Lemma 2} to the top-right grid (i.e., $\theta_{1,2},\theta_{1,3},\theta_{2,2},\theta_{2,3}$), let $\sin(\theta_{1,3}-\theta_{1,2}) = x$ and $\rho = \frac{\omega}{n^2k}$, in which $n = 3$ in this case. Taking $\delta_1 = x, \delta_2 = x+\rho, \delta_3 = \frac{5\rho}{2}+x,\delta_4 = 3\rho+x$, we can find that the cosine of the phase shift vector $\begin{bmatrix}\cos(\phi_1)\ \cos(\phi_2)\ \cos(\phi_3)\ \cos(\phi_4)\ \end{bmatrix}$ is parallel to:

\begin{align}
    \begin{bmatrix}
    2(13x+20\rho) \\
    -2(11x+19\rho)\\
    -(14x+25\rho)\\
    10x+14\rho
    \end{bmatrix}
\end{align}

By calculating $\frac{\cos(\phi_1)}{\cos(\phi_4)} = \frac{13x+20}{5x+7}$ and $\frac{\cos(\phi_2)}{\cos(\phi_4)} = \frac{11x+19}{5x+7}$, the equation that $x$ needs to satisfy is obtained:

\be
  24x^3+208\rho x^2+8(65\rho^2-3)x+(400\rho^2-39)\rho = 0
\ee

Since $x$ is solved, all phase shifts are also obtained simultaneously, which is consistent to the numerical simulation in Fig. \ref{fig:resultIII21}. 

\subsubsection{Wave propagation in the $n \times n$ grid with noise}

Next, consider a larger grid and let all oscillators rotate from the zero phase. Lemma 1 shows that for the "driver" oscillator $\theta_{1,1}$, the equilibrium state of phase shifts is:

\be \label{eq: maximum equilibrium condition}
 \sin(\theta_{1,1}-\theta_{1,2}) =\sin(\theta_{1,1}-\theta_{2,1}) = \frac{(n^2-1)}{2}\rho
\ee

For the "passenger" oscillator $\theta_{n,n}$, the equilibrium state of phase shifts $\sin(\theta_{n-1,n}-\theta_{n,n}) = \sin(\theta_{n,n-1}-\theta_{n,n})$ equals $\frac{1}{2}\rho$. For oscillators between these two, their phase shifts between each other are also located in this interval, and this wave propagates along the direction of decreasing $\sin(\phi_{ij})$. 

Therefore, both $\omega$ and $k$ can determine the wave pattern: (1) If $\omega$ increases, the average speed $\bar{\omega}$ also increases, as well as phase shifts. (2) If $k$ increases, the phase shifts will decrease which means the all oscillators synchronize more closely. 

Now consider the general case to explore the influence of noise. Assume that intrinsic frequencies of all oscillators are taken from a normal distribution $\mathcal{N}(\omega_l,\sigma_{\omega})$ , except for "driver oscillator" which is taken from the distribution $\mathcal{N}(\omega_{h}, \sigma_{\omega}^2)$ with a larger intrinsic frequency, i.e., $\omega_{h}>\omega_{l}$ and define $\omega_h-\omega_l = \omega$. Denote $\Delta \omega_i  = \omega_i - \omega_l, i = 2,..,n^2$ and $\Delta \omega_1 = \omega_1 - \omega_h,$. Assume that coupling strengths $k_{ij}$ are taken from $\mathcal{N}(\mu_k,\sigma_k^2)$. To simplify the problem, let $\sigma_k \ll \mu_k$. By \textbf{Lemma 1}, the condition of equilibrium state is determined by $\frac{d(\theta_i -\bar{\omega}t)}{dt}$, or:

\be
\sum_{j \in \Lambda_i} k_{ji}\sin (\theta_j - \theta_i) =\bar{\omega}-\omega_i
\ee

If the noise $\sigma_l$ is close to zero, then it is similar to the simple case discussed before. Therefore, it is natural to introduce a linear perturbation $\Delta \theta_i = \theta_i - \theta_{i}^*$ on each oscillator when the noise increase. Here $\theta_i^*$ satisfy:

\be \label{eq:equilibrium condition}
\sum_{j \in \Lambda_i} k_{ji}\sin (\theta_j^* - \theta_i^*) =\frac{\omega_h-\omega_l}{n^2}+(\omega_l-\omega_i)
\ee

Taking the first order expansion by $\sin(x+\Delta x) \approx \sin(x) + \cos(x)\Delta x +o(\Delta x^2)$, the new condition of equilibrium state becomes:

\begin{equation}
 \begin{aligned}
 \sum_{j \in \Lambda_i} k_{ji}\sin (\theta_j^* &- \theta_i^*) + \sum_{j \in \Lambda_i} k _{ji}\cos (\theta_j^* - \theta_i^*)(\Delta \theta_j-\Delta \theta_i)\\
 &=\frac{\omega_h-\omega_l}{n^2}+(\omega_l-\omega_i)-\frac{\sum \Delta \omega_i}{n^2}-\Delta \omega_i
 \end{aligned}
\end{equation}

With eq.\ref{eq:equilibrium condition}, above equations are reduced to:

\be \label{eq:error equilibrium condition}
\sum_{j \in \Lambda_i} k _{ji}\cos (\theta_j^* - \theta_i^*)(\Delta \theta_j-\Delta \theta_i)=-\frac{\sum \Delta \omega_i}{n^2}-\Delta \omega_i
\ee

Since $\Delta \omega_i \sim \mathcal{N}(0, \sigma_{\omega}^2)$, then $\frac{\sum \Delta \omega_i}{n^2}$ is close to $0$ when $n$ is large. On the other hand $\cos(\theta^*_j - \theta_i^*) = \sqrt{1-\sin(\theta^*_j - \theta_i^*)^2} \geq \sqrt{1-\frac{\sigma^2}{\mu_k^2}\frac{(n^2-1)^2}{(2n^2)^2}}\geq \sqrt{1-\frac{\sigma^2}{4\mu_k^2}}$ and most of phase difference are much little so that the $\cos$ terms are close to 1. By dividing $\mu_k$ on both sides of eq.\ref{eq:reduced error equilibrium condition} and denoting $\widetilde{k_{ij}} = \frac{k_{ij}}{\mu_k}, \widetilde{\Delta \omega_i} = \frac{\Delta \omega_i}{\mu_k}$,
an approximation of eq.\ref{eq:error equilibrium condition} is:

\be \label{eq:reduced error equilibrium condition}
 \sum_{j \in \Lambda_i} \widetilde{k_{ji}} (\Delta \theta_j-\Delta \theta_i)=-\widetilde{\Delta \omega_i} 
\ee

By assumption $\sigma_k \ll \mu_k$, therefore $\widetilde{k_{ij}} \sim \mathcal{N}(1, \frac{\sigma^2_k}{\mu_k^2})$ is around 1. On the other hand, $\widetilde{\Delta \omega_i} \sim \mathcal{N}(0,\frac{\sigma_\omega^2}{\mu_k^2})$. 

To measure how the noise perturb the wave propagation, here provides an index,

\be
 r_{\omega} = \frac{\theta_{1,1} - \theta_{n,n}}{\max \{\theta_{i,j}-\theta_{n,n}\mid i,j\in\{1,2,\dots,n\}\}}
\ee

Intuitively, the index is $1$ when there is no noise since $\theta_{1,1}$ is the "driver" oscillator and wave propagates to the end oscillator $\theta_{n,n}$. However, when the noise increases, the leadership of $\theta_{1,1}$ may be sabotaged by other oscillators, i.e., the intrinsic frequency of one oscillator is extreme large due to huge $\sigma_\omega$ and even exceed the "driver", then the index $ r_{\omega}$ will below 1. By keeping $\mu_k$ and change $\sigma_\omega$, for example $\mu_k = 4$, the index $ r_{\omega}$ will remains at $1$ at first and decrease soonly after $\sigma_\omega$ exceeds $0.1$.  Simulation result implies that the wave propagation direction will remain under a small noise but vanish soonly if increasing the noise.  (Red line in Fig.\Ref{fig:resultnoise})

Although the analytic form of $ r_{\omega} $ is hard to get, an approximation form can be addressed based on eq.\ref{eq:reduced error equilibrium condition}. Replace $\theta$ with $\theta^*+\Delta \theta$, then $\theta_{1,1}-\theta_{n,n} = \theta_{1,1}^*-\theta_{n,n}^*+\Delta \theta_{1,1} - \Delta \theta_{n,n}$ and $\max\{\theta_{i,j}-\theta_{n,n}\} = \theta_{1,1}^*-\theta_{n,n}^* + \Delta\theta_{1,1}-\Delta\theta_{n,n} + \max\{  \theta_{i,j}-\theta_{1,1} \}$. For a given $n$, $\theta_{1,1}^*-\theta_{n,n}^*$ is a fixed value. Denote the variance of $\Delta \theta_i$ is $\widetilde{\sigma_\omega}$ so that the scale of $ \Delta\theta_{1,1}-\Delta\theta_{n,n}$ is roughly governed by the distribution $\mathcal{N} (0, 2\widetilde{\sigma_\omega}^2 )$. For $\max\{  \theta_{i,j}-\theta_{1,1} \}$, if noise is small then the oscillator $ \{i,j\}$ would be close to the "driver" and this difference is close to $0$. If noise is large then the difference $\theta^*_{i,j}-\theta^*_{1,1}$ would be overwhelmed by the noise, in which is governed by the expectation of extreme (maximal) value distribution of $\mathcal{N} (0, 2\widetilde{\sigma_\omega}^2 )$. In all, an approximation form $\widetilde{r_{\omega}}$ is:

\be \label{eq:approximation of r}
 \widetilde{r_{\omega}} = \frac{\theta_{1,1}^*-\theta_{n,n}^*+ \sqrt{2}\widetilde{\sigma_\omega}}{\theta_{1,1}^*-\theta_{n,n}^*+ \sqrt{2}\widetilde{\sigma_\omega}+ \sqrt{2}\widetilde{\sigma_\omega}\Phi^{-1}(\frac{n^2-\frac{\pi}{8}}{n^2-\frac{\pi}{4}+1})}
\ee

\begin{figure}[htbp]
    \centering
    \includegraphics[width=0.75\textwidth]{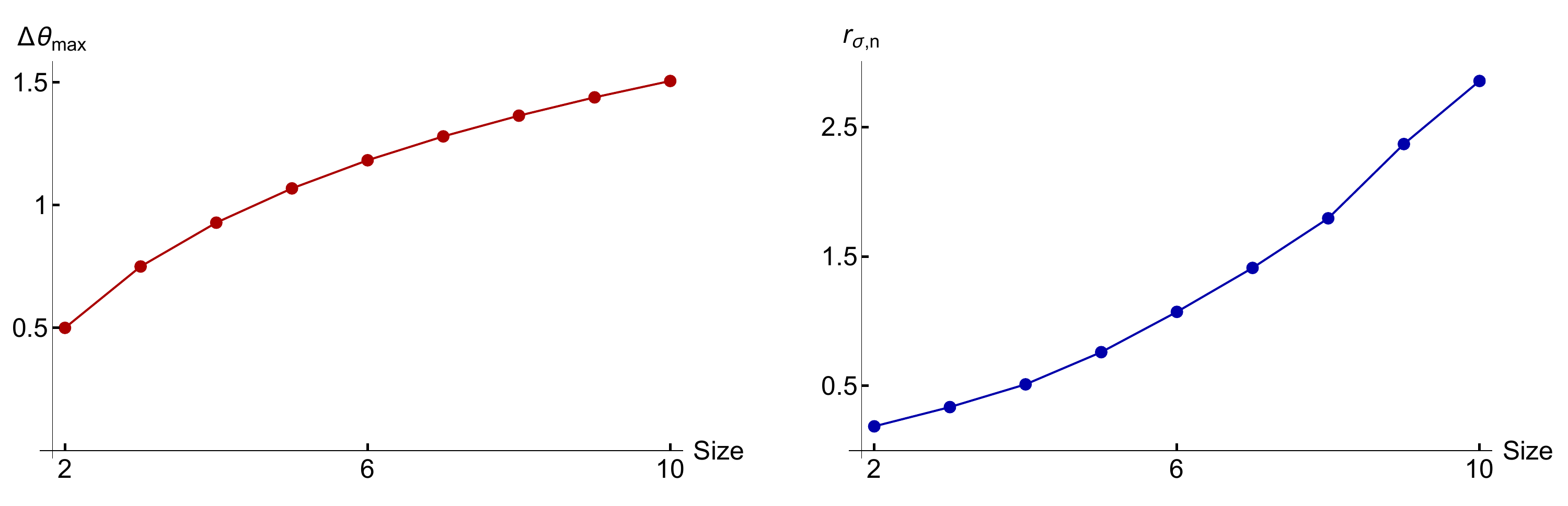}
    \caption{Left: $\theta_{1,1}^*-\theta^*_{n,n}$. Right: Mean of $r_{\sigma,n}$ when $\mu_k = 1$.}
    \label{fig:thetarn}
  \end{figure}

\begin{figure}[htbp]
    \centering
    \includegraphics[width=0.5\textwidth]{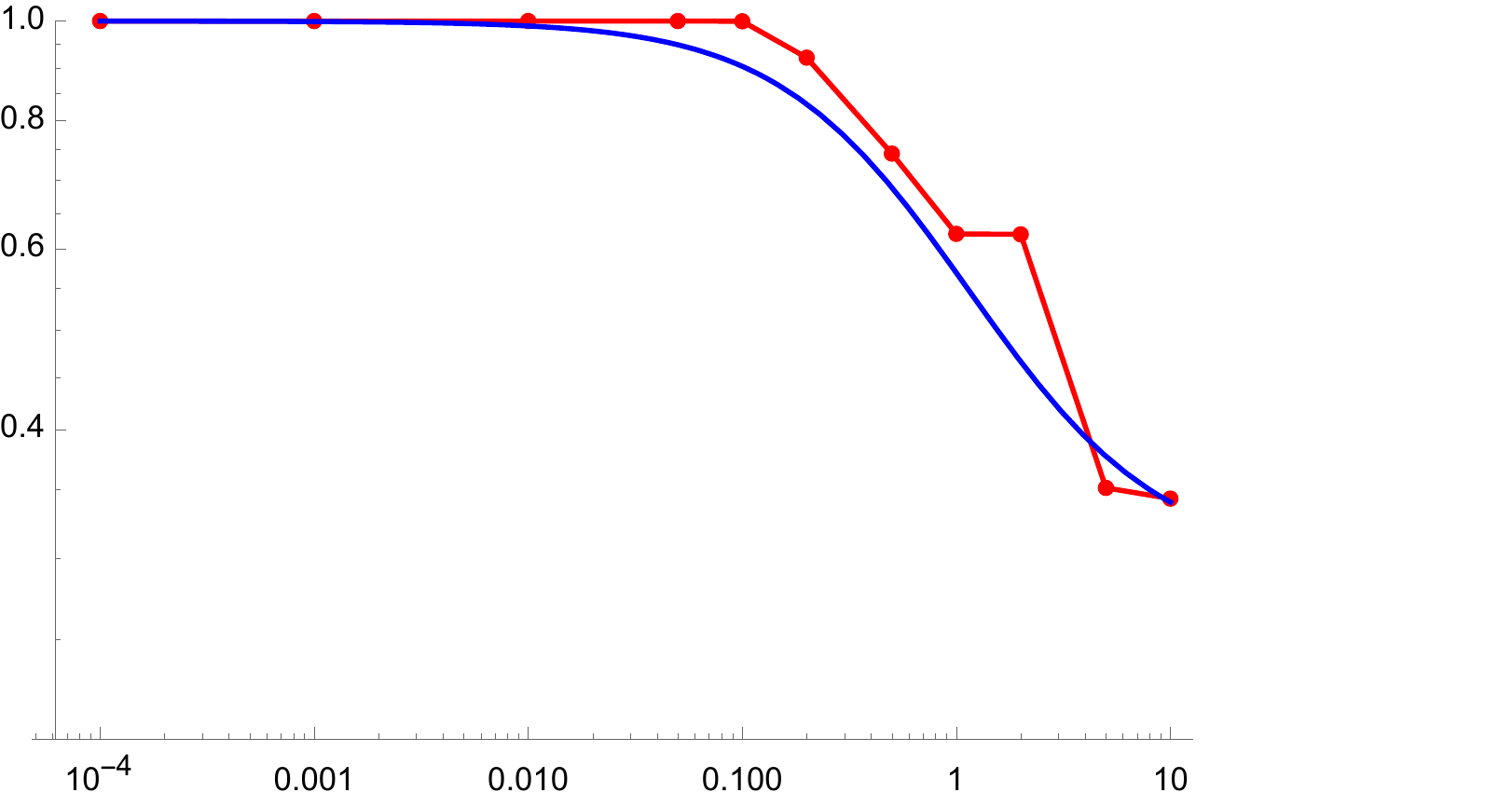}
    \caption{Directional wave propagation vanishes with increasing noise. Red line: Numerical simulation. Each points contains 400 trials. Simulation parameters: $n = 8, \omega_{\text{fast}} \sim \mathcal{N}(2, \sigma), \omega_{\text{slow}} \sim \mathcal{N}(1, \sigma), k_{ij} \sim \mathcal{N}(4, 0.05)$, and $\sigma = 0.0001,0.001,0.01,0.1,0.2,0.4,0.8,1,2,4,10$. Blue line: Prediction by eq.\ref{eq:approximation of rw}.}
    \label{fig:resultnoise}
  \end{figure}

In eq.\ref{eq:approximation of r}, the explicit form of $\theta_{1,1}^*-\theta^*_{n,n}$ is hard to solve, and the exact values with no noise for the beginning ten can be found in the left of Fig.\Ref{fig:thetarn}. Finding $\widetilde{\sigma_{\omega}}^2$ is also a troublesome task, and it can be numerically observed that there is a proportion $ r_{\sigma,n} $ between $\widetilde{\sigma_{\omega}}^2$ and $\frac{\sigma_\omega^2}{\mu_k^2}$ that increases with $n$, which is shown in the right of Fig.\Ref{fig:thetarn}. Therefore $\widetilde{r_{\omega}}$ can also expressed as a function of $\frac{\sigma_{\omega}}{\mu_k}$:

\be \label{eq:approximation of rw}
 \widetilde{r_{\omega}}(\sigma_\omega) = \frac{\theta_{1,1}^*-\theta_{n,n}^*+ \sqrt{2 r_{\sigma,n} }\frac{\sigma_\omega}{\mu_k}}{\theta_{1,1}^*-\theta_{n,n}^*+ \sqrt{2 r_{\sigma,n} }\frac{\sigma_\omega}{\mu_k}+ \sqrt{2 r_{\sigma,n} }\frac{\sigma_\omega}{\mu_k}\Phi^{-1}(\frac{n^2-\frac{\pi}{8}}{n^2-\frac{\pi}{4}+1})}
\ee

Eq.\ref{eq:approximation of rw} qualitatively shows how the increasing noise $\sigma_\omega$ destroys the wave propagation and how the increasing coupling strength $\mu_k$ slows down this destruction. See FIG.\Ref{fig:resultnoise}. More importantly, if the system does not have an significantly fast oscillator that can overcome the noise, a directional wave propagation is hard to form. 

\subsubsection{Overspeed case}

Finally, spend some space explaining the state of the system when the equilibrium condition cannot be satisfied, that is, when the equilibrium condition eq.\ref{eq: maximum equilibrium condition} cannot be met, which is equivalent to $\frac{\omega}{\mu_k}> \frac{2n^2}{n^{2}-1}$. Either too large intrinsic frequency of "driver" or a weak coupling network would cause this problem, but in different ways. 

For the former, the phase difference between the “driver” and other oscillators in the system will accumulate continuously. When $\omega$ is sufficiently large, the constraints imposed by the coupling term will have no effect, and the plasticity of the accumulated phase difference will be determined by $\omega$.

For the latter, if the coupling strength is too weak, the "driver" may not be able to effectively accelerate even the surrounding oscillators. If all oscillators in the system start from zero phase, the order parameter will first decrease and then increase with the increase of the coupling strength. This turning point is coincident with $k^* =  \frac{n^2-1}{2n^2}\omega$, which implies the establishment of wave propagation in the system. 

\begin{figure}[htbp]
    \centering
    \includegraphics[width=0.5\textwidth]{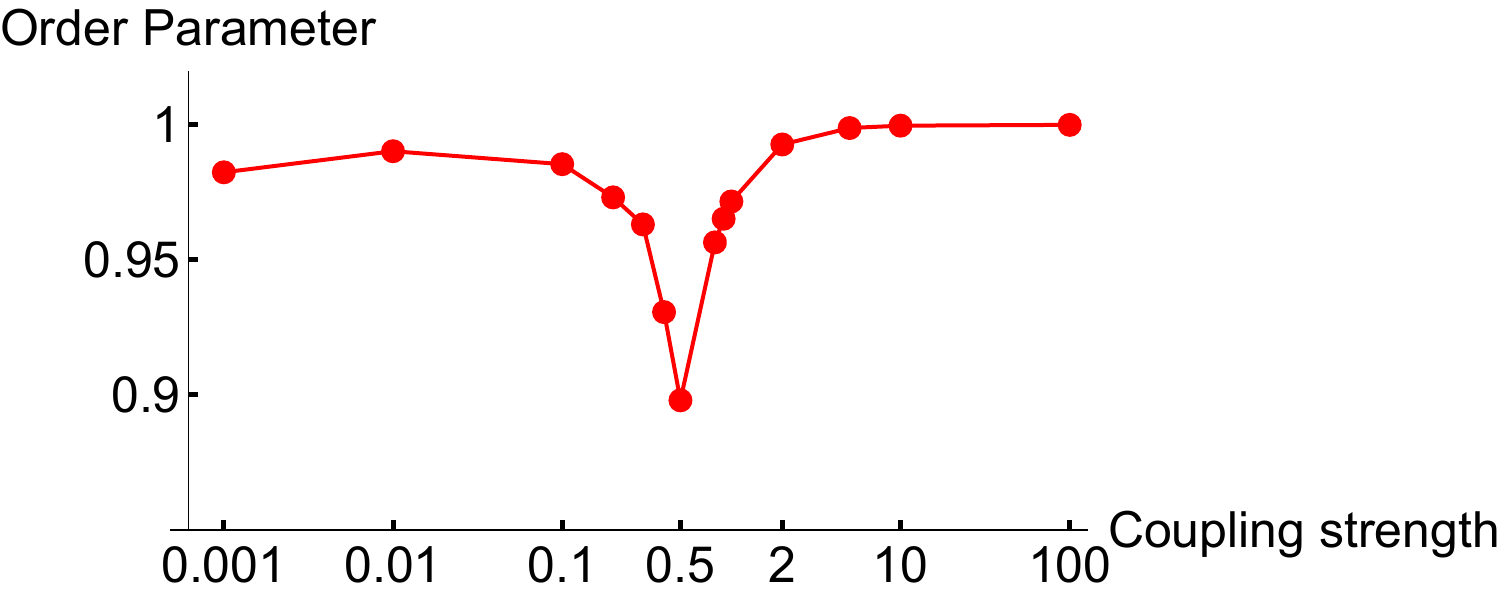}
    \caption{Order parameter nonmonotonically changes with increasing coupling strength. In this case, $n = 8$, $\omega = 1$.}
    \label{fig:small k}
  \end{figure}

\subsection{Spiral wave}

In the last section, I discussed how the directional wave propagation forms, with a stronge assumption that all oscillators start from zero phase. This assumption does not conform to the general practice of taking initial values, which is to randomly select an initial phase from the interval $[0,2\pi]$. Previous numerical simulation results have also shown that for systems with the same parameters, if the initial phase value distribution is one in $[0,\pi]$ and the other in $[0,2\pi]$, the latter is more likely to generate spiral waves at some places. In this section, I will start from a $4\times 4$ grid to discuss the causes, analytical solutions, and stability of spiral wave patterns.

In the previous example of the degenerate $2\times2$ grid, two types of spiral wave patterns were shown to exist, but neither was stable. In this section, I will analytically derive the conditions for the phase differences that give rise to spiral wave patterns, starting from the quadrilateral rotational symmetry, and it can be shown that this pattern exists stably.

\begin{figure}[htbp]
  \centering
  \includegraphics[width=0.5\textwidth]{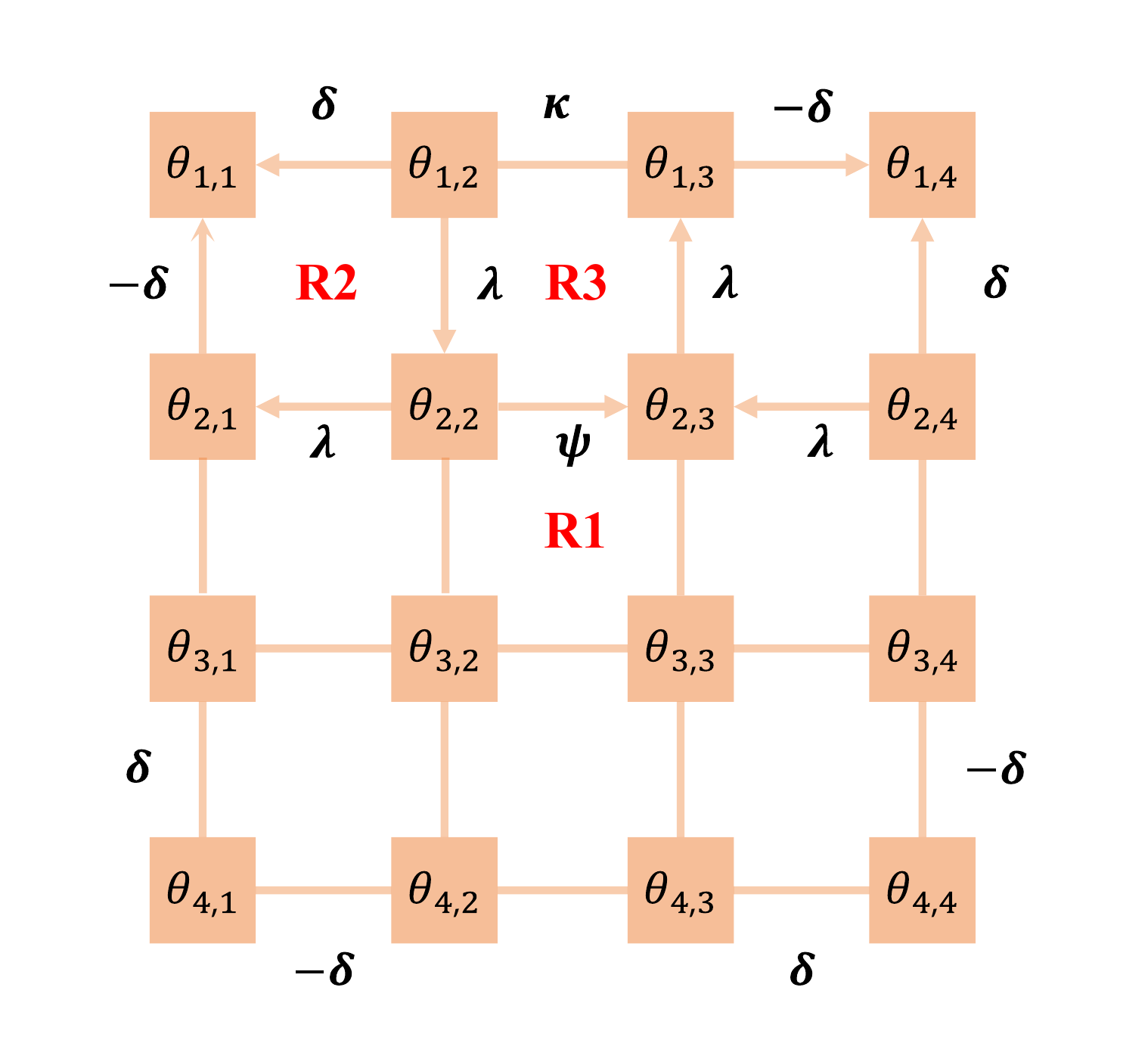}
  \caption{Oscillators on a $4 \times 4$ grid}
  \label{fig 4 by 4 grid}
\end{figure}

\subsubsection{ Analytic solution to $4\times 4$ grid}
The oscillators are arranged and named as shown in the figure for a $4\times4$ grid, and it is assumed that all oscillators have the same intrinsic frequency $\omega$ and coupling strength $k_{ij} = \mu_k$. First, notice that the oscillator at the top-left corner $\theta_{1,1}$ satisfies the condition $\sin(\theta_{1,2}-\theta_{1,1})+\sin(\theta_{2,1}-\theta_{1,1}) = 0$ so that denote $\sin(\theta_{1,2}-\theta_{1,1}) = \sin(\theta_{1,1}-\theta_{2,1}) = \delta$. Due to the quadrilateral rotational symmetry, the oscillators at four corners also follows that 

\begin{align}
    \sin(\theta_{1,2}-\theta_{1,1}) &=\sin( \theta_{2,4}-\theta_{1,4}) \notag\\
    &=\sin( \theta_{4,3}-\theta_{4,4})\\
    &=\sin( \theta_{3,1}-\theta_{4,1}) = \delta \notag\\
    \sin(\theta_{2,1}-\theta_{1,1}) &=\sin( \theta_{1,3}-\theta_{1,4}) \notag\\
    &=\sin( \theta_{3,4}-\theta_{4,4})\\
    &=\sin( \theta_{4,2}-\theta_{4,1}) = -\delta \notag
\end{align}

Next, use the equilibrium conditions on $\theta_{1,2}$ and $\theta_{1,3}$ and denote $\sin(\theta_{1,2}-\theta_{2,2}) = \lambda_1, \sin(\theta_{2,3}-\theta_{1,3}) = \lambda_2, \sin(\theta_{1,3}-\theta_{1,2}) = \kappa$, then:

\begin{equation}
 \begin{aligned}
 -\delta + \kappa - \lambda_1 &= 0\\
 \delta - \kappa + \lambda_2 &= 0
 \end{aligned}
\end{equation}

It gives $-\lambda_1 + \lambda_2 = 0$. Denote $\lambda_1 = \lambda_2 = \lambda$, therefore the following relations are true due to the quadrilateral rotational symmetry:

\begin{align}
    \sin(\theta_{1,2}-\theta_{2,2}) &=\sin( \theta_{2,4}-\theta_{2,3}) \notag\\
    &=\sin( \theta_{4,3}-\theta_{3,3})\\
    &=\sin( \theta_{3,1}-\theta_{3,2}) = \lambda \notag\\
    \sin(\theta_{2,3}-\theta_{1,3}) &=\sin( \theta_{3,3}-\theta_{3,4}) \notag\\
    &=\sin( \theta_{3,2}-\theta_{4,2})\\
    &=\sin( \theta_{2,2}-\theta_{2,1}) = \lambda \notag
\end{align}

For a oscillator in the core, i.e., $\theta_{2,2}$, it follows $\sin(\theta_{1,2}-\theta_{2,2})+\sin(\theta_{2,1}-\theta_{2,2})+\sin(\theta_{2,3}-\theta_{2,2})+\sin(\theta_{3,2}-\theta_{2,2}) = 0$. Since $\sin(\theta_{1,2}-\theta_{2,2})+\sin(\theta_{2,1}-\theta_{2,2}) = 0$, it induced $\sin(\theta_{2,2}-\theta_{2,3}) = \sin(\theta_{3,2}-\theta_{2,2})$, denoting as $\psi$. Therefore,

\begin{equation}
 \begin{aligned}
  \sin(\theta_{2,2}-\theta_{2,3}) &= \sin(\theta_{3,2}-\theta_{2,2}) \\
  &= \sin(\theta_{3,3}-\theta_{3,2}) \\
  &= \sin(\theta_{2,3}-\theta_{3,3}) = \psi
 \end{aligned}
\end{equation}

The equilibrium states of all phase difference are settled, which can be represented by variables $\delta, \kappa, \lambda, \psi$. Using \textbf{Lemma 2}, we can further establish a connection between these variables. See FIG. \ref{fig 4 by 4 grid}. 

\vspace*{5pt}
\textbf{R1}. For the central $2\times2$ grid, its equilibrium state is $\delta_1=\delta_2=\delta_3=\delta_4 = \psi$, which is the same as the degenerate $2\times2$ grid discussed earlier. Therefore, there are also three possible values for $\psi$, among which the first type is a trivial solution and will not be discussed further. We focus on the second and third types of solutions, which differ in whether $\cos(\theta_i)$ are equal. If they are equal, the solution corresponds to a quadrilateral rotation symmetry, while if they differ by a negative sign pairwise, the solution corresponds to a bilateral-like rotation symmetry. 

\vspace*{5pt}
\textbf{R2}. For the $2\times 2$ grids in the corners, take the top left one $\theta_{1,1},\theta_{1,2},\theta_{2,2},\theta_{2,1}$ as an example, its equilibrium state is $\sin(\theta_{1,2}-\theta_{1,1}) = \sin(\theta_{1,1}-\theta_{2,1}) = \delta, \sin(\theta_{2,2}-\theta_{1,2})= \sin(\theta_{2,2}-\theta_{2,1}) = \lambda$. Due to the identical equation $(\theta_{1,2}-\theta_{1,1})+(\theta_{2,2}-\theta_{1,2})+(\theta_{2,1}-\theta_{2,2}) = -(\theta_{1,1}-\theta_{2,1})$, using the same notation in eq. \ref{eq:def} and taking sine on both sides:

\begin{equation} \label{eq. threeone}
 \begin{aligned}
 &\sin(\phi_1)\cos(\phi_2)\cos(\phi_3) +\sin(\phi_2) \cos(\phi_1)\cos(\phi_3) \\
 &+ \sin (\phi_3) \cos(\phi_1) \cos(\phi_2) - \sin(\phi_1)\sin(\phi_2)\sin(\phi_3) \\
 &= -\sin(\phi_4)
 \end{aligned}
\end{equation} 

Denote $\cos(\phi_1) =\delta^*$. Since $\sin(\phi_2) = \sin(\phi_3)$, the cosine values are either identical or opposite:

\textbf{a.} $\cos(\phi_2) = \cos(\phi_3) = \lambda^*$, then eq.\ref{eq. threeone} becomes:
\be
 \delta (\lambda^*) ^2 + 2\lambda\lambda^*\delta^* -\lambda^2\delta = -\delta
\ee

Replacing $\lambda^2 $ by $ 1 -(\lambda^*)^2$, then:
\begin{equation}\label{eq:R2}
 \lambda \delta^* + \lambda^* \delta  = 0 \iff \sin(\phi_1+\phi_2) = 0
\end{equation}

\vspace*{3pt}
If $\sin(\phi_1) = -\sin(\phi_2)$, it gives the trivial solution.

\textbf{b.} $\cos(\phi_2) = -\cos(\phi_3) = \lambda^*$, then eq.\ref{eq. threeone} gives an indetical equation:

\be
 (\lambda^*)^2 + \lambda^2 = 1
\ee

In this scenario, $\lambda$ do not need to be indetical to $\delta$. 

\vspace*{5pt}
\textbf{R3}. For the $2\times 2$ grids located on the edge, such as $\theta_{1,2},\theta_{1,3},\theta_{2,3},\theta_{2,2}$, the equilibrium state is $\delta_1 = \delta + \lambda,\delta_2 = \delta_4 = \lambda,\delta_3 = \psi$.  \textbf{Lemma 2} indicates that the cosine of the phase shift vector is parallel to:

\begin{align} \label{eq:r3}
 \begin{bmatrix} 
    \delta^2 + 2 \delta \lambda - \lambda^2 + \psi (\delta + \lambda)\\ 
    \lambda (\psi - \delta - \lambda)\\
    2\lambda^2 - \psi(\delta+\lambda)-\psi^2\\
    \lambda (\psi - \delta - \lambda)
\end{bmatrix} 
\end{align}

According to \textbf{R2}, we can classify the wave pattern based on whether $\delta$ equals $\lambda$ or not:

\textbf{W1. $\delta = \lambda$}.  
\vspace*{5pt}

By eq.\ref{eq:r3}, 

\begin{equation} \label{eq:r3 s}
 \begin{aligned}
 \frac{\cos(\theta_{2,2}-\theta_{2,3})}{\cos(\theta_{2,3}-\theta_{1,3})} 
 &= \frac{\cos(\theta_{2,3}-\theta_{3,3})}{\cos(\theta_{3,3}-\theta_{3,4})}\\
 &= \frac{\cos(\theta_{3,3}-\theta_{3,2})}{\cos(\theta_{3,2}-\theta_{4,2})}\\
 &= \frac{\cos(\theta_{3,2}-\theta_{2,2})}{\cos(\theta_{2,2}-\theta_{2,1})}\\
 &= \frac{2\lambda^2-2\lambda \psi -\psi^2}{\lambda(\psi-2\lambda)}
 \end{aligned}
\end{equation}

And by \textbf{R2.a},

\begin{equation} \label{eq:r2 s}
    \begin{aligned}
    \frac{\cos(\theta_{2,1}-\theta_{2,2})}{\cos(\theta_{2,2}-\theta_{1,2})} 
    &= \frac{\cos(\theta_{1,3}-\theta_{2,3})}{\cos(\theta_{2,3}-\theta_{2,4})}\\
    &= \frac{\cos(\theta_{3,3}-\theta_{3,3})}{\cos(\theta_{3,3}-\theta_{3,4})}\\
    &= \frac{\cos(\theta_{4,2}-\theta_{3,2})}{\cos(\theta_{3,2}-\theta_{3,1})}\\
    &= 1
    \end{aligned}
\end{equation}

It concludes that the cosine of the phase shifts in core are indetical:

\begin{equation}
    \begin{aligned}
        \cos(\theta_{2,2}-\theta_{2,3}) &= \cos(\theta_{2,3}-\theta_{3,3}) \\
        &= \cos(\theta_{3,3}-\theta_{3,2}) \\
        &= \cos(\theta_{3,2}-\theta_{2,2})
    \end{aligned}
\end{equation}

\begin{figure}[htbp]
    \centering
    \includegraphics[width=0.8\textwidth]{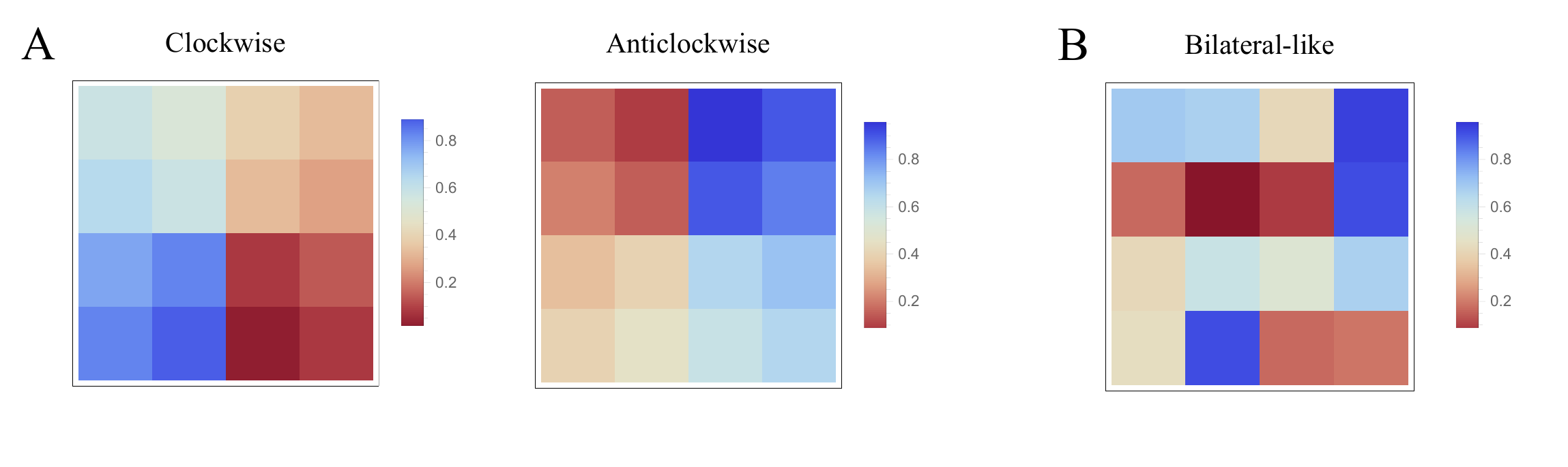}
    \caption{Wave pattern of \textbf{W1} and \textbf{W2}. A: Wave pattern of \textbf{W1}. B: Wave pattern of \textbf{W2}.}
    \label{fig:resultIV1}
  \end{figure}

In this case, $\psi = 1, \delta = \lambda = \frac{-1+\sqrt{3}}{2}$ (\textbf{Clockwise}), or $\psi = -1, \delta = \lambda = \frac{1-\sqrt{3}}{2}$ (\textbf{Anti-clockwise}). See FIG. \ref{fig:resultIV1}A.

\vspace*{5pt}

\textbf{W2. $\delta \neq \lambda$}.  
\vspace*{5pt}

Eq.\ref{eq:r3 s} is still true in this scenario, but eq.\ref{eq:r2 s} becomes:

\begin{equation}
    \begin{aligned}
    \frac{\cos(\theta_{2,1}-\theta_{2,2})}{\cos(\theta_{2,2}-\theta_{1,2})} 
    &= \frac{\cos(\theta_{1,3}-\theta_{2,3})}{\cos(\theta_{2,3}-\theta_{2,4})}\\
    &= \frac{\cos(\theta_{3,3}-\theta_{3,3})}{\cos(\theta_{3,3}-\theta_{3,4})}\\
    &= \frac{\cos(\theta_{4,2}-\theta_{3,2})}{\cos(\theta_{3,2}-\theta_{3,1})}\\
    &= -1
    \end{aligned}
\end{equation}

And the cosine of the neighboring phase shifts in core are opposite:

\begin{equation}
    \begin{aligned}
        \cos(\theta_{2,2}-\theta_{2,3}) &= -\cos(\theta_{2,3}-\theta_{3,3}) \\
        &= \cos(\theta_{3,3}-\theta_{3,2}) \\
        &= -\cos(\theta_{3,2}-\theta_{2,2})
    \end{aligned}
\end{equation}

Therefore $\psi$ is not a determined value, as well as $\lambda$. Here is an example, let $ \theta_0 = \arcsin(\frac{2-\sqrt{3}}{2}) $, then following intital condition matrix $\Theta_0$ generates the wave pattern predicted in \textbf{W2.}: 

\begin{align}
   \Theta_0 =  \begin{bmatrix} 
        \theta_0 + \frac{4\pi}{3} &  \frac{4\pi}{3}  &  \frac{5\pi}{6} &   \theta_0 + \frac{11\pi}{6}\\
        \frac{\pi}{3} & 0 & \frac{\pi}{6} & \frac{11\pi}{6} \\
        \frac{5\pi}{6} & \frac{7\pi}{6} & \pi & \frac{4\pi}{3} \\
        \theta_0 + \frac{5\pi}{6} &  \frac{11\pi}{6}  &  \frac{\pi}{3} &   \theta_0 + \frac{\pi}{3}
    \end{bmatrix} 
\end{align}       

In this case, $\psi = -\frac{1}{2}, \lambda = -\frac{\sqrt{3}}{2}, \delta = \frac{-2+\sqrt{3}}{2}, \kappa = -1$. See FIG. \ref{fig:resultIV1}B.

\subsubsection{Stability of W1 and W2}

In section \textbf{III}, I have discussed the stability of \textbf{S1} and \textbf{S2} by calculating the eigenvalues of their Jacobian matrix, in which \textbf{S1} is in central manifold while \textbf{S2} is an saddle point. Since \textbf{W1} and \textbf{W2} are developed based on \textbf{S1} and \textbf{S2}, respectively, their stability properties are also similar. 

The numerical solution of \textbf{S1} has been solved, the numerical solutions of the eigenvalues of the corresponding Jacobian matrix can also be obtained, where the largest eigenvalue ($\approx -0.22502$) is also less than 0. It concludes that \textbf{W1} is a stable wave pattern.

As for \textbf{W2}, there are more than one possible numerical solutions, so the determination of whether the eigenvalues are greater than 0 can only be based on how they depend on the parameters. Fortunately, although the form of the eigenvalue system of this matrix is complex, a pair of eigenvalues can be obtained analytically:

\begin{equation}
 \begin{aligned}
    \lambda_1 = -\kappa^* - \sqrt{3(\delta^*)^2 - 2\delta^*\lambda^* + 3(\lambda^*)^2 + (\kappa^*)^2  } \\
    \lambda_2 = -\kappa^* + \sqrt{3(\delta^*)^2 - 2\delta^*\lambda^* + 3(\lambda^*)^2 + (\kappa^*)^2  } 
 \end{aligned}
\end{equation}

\begin{figure}[htbp]
    \centering
    \includegraphics[width=0.6\textwidth]{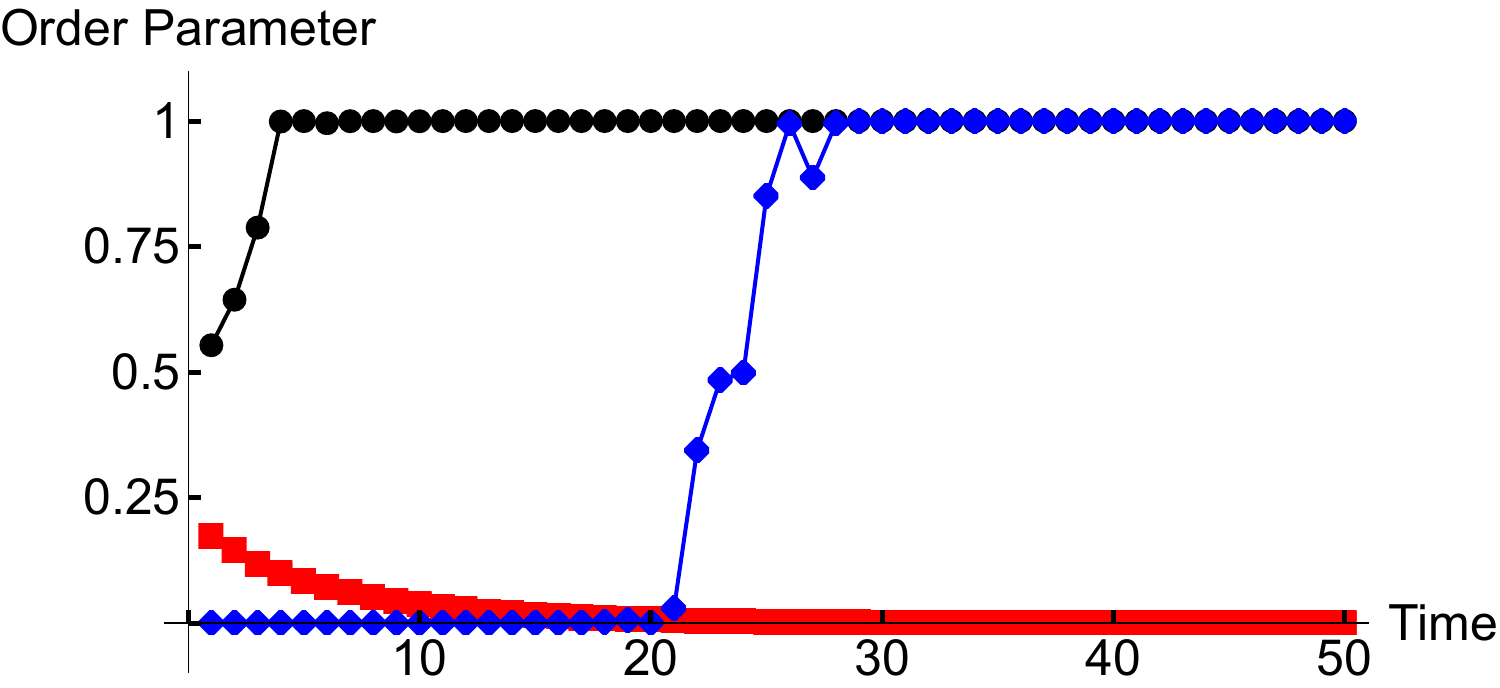}
    \caption{Time traces of order parameter starting with different initial conditions. Black: Start from a random initial condition. Red: Start from a initial condition that converges to \textbf{W1}. Blue: Start from \textbf{W2}. Simulation parameters: $n = 4, \omega_i = 1, k_{ij} = 0.1$.}
    \label{fig:resultIV3}
  \end{figure}

Here, $\delta^*, \lambda^*, \kappa^*$ are the corresponding cosine value of $\delta, \lambda, \kappa$, i.e., their sum of squares is 1. Notice that $\lambda_2 = \sqrt{(\delta^*-\lambda^*)^2+2(\delta^*)^2 +2(\lambda^*)^2 + (\kappa^*)^2 } - \kappa^*  >0 $ is always true, the dimension of unstable manifold is at least $1$. It is also consistent to the observation in numerical simulation that even the exact initial condition is given, \textbf{W2} pattern still crashes after several iteration steps. (FIG. \ref{fig:resultIV3})

In all, it was found that \textbf{W0} and \textbf{W1} are stable solutions, as their eigenvalues are both less than 0. On the other hand, \textbf{W2} is a first-order saddle point, as it only has one eigenvalue greater than 0.

\subsubsection{Spiral wave pattern in the $n \times n$ grid}

\begin{figure}[htbp]
  \centering
  \includegraphics[width=0.8\textwidth]{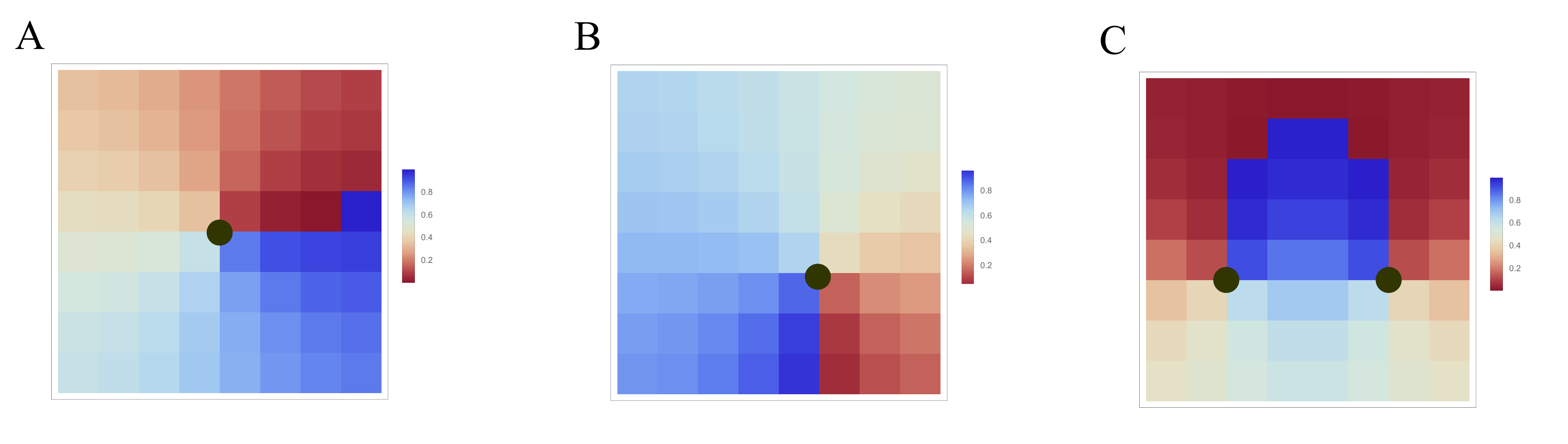}
  \caption{Spiral wave patterns in a larger grid, i.e. $8\times 8$ grid. A. Spiral wave in the center. B. Spiral wave bias from center. C. Two spiral waves coexist.Simulation parameters: $n = 8, \omega_i = 1, k_{ij} = 1$.}
  \label{fig:resultIV2_2 mul}
\end{figure}

Although analytical solutions and stability conditions have been obtained for spiral waves on a $4\times 4$ lattice, the situation becomes more complex as the system size $n$ exceeds 4. Based on the result of \textbf{W1}, a simple speculation for an $n\times n$ grid is that it can also form a stable spiral wave in the center. Based on \textbf{R2}, a simple inference is that for each small square on the diagonal, the cosine values of the phase differences between adjacent oscillators are equal, and the coupling term has quadrilateral rotational symmetry overall, i.e., as shown in FIG. \ref{fig:resultIV2_2 mul}A.

However, with increasing lattice size, numerical simulations show two variations of spiral waves. Firstly, The rotation center can appear in positions far away from the central grid point. These "eccentric" spiral waves do not have strict quadrilateral rotational symmetry like the "standard" spiral wave located at the center. In fact, they may not even have rotation symmetry in a strict sense, due to the finite number of grid points in the model and the lack of periodic boundary conditions, which means that all grid points except the center do not have good rotational symmetry. Intuitively, when there are enough grid points to locally accommodate a $4\times4$ square, it is possible to understand the generation of an approximate rotationally symmetric solution in this local area. Here is the example spiral wave in a $8 \times 8$ grid which rotation center locates besides center. (FIG. \ref{fig:resultIV2_2 mul}B)

\begin{figure}[htbp]
  \centering
  \includegraphics[width=0.8\textwidth]{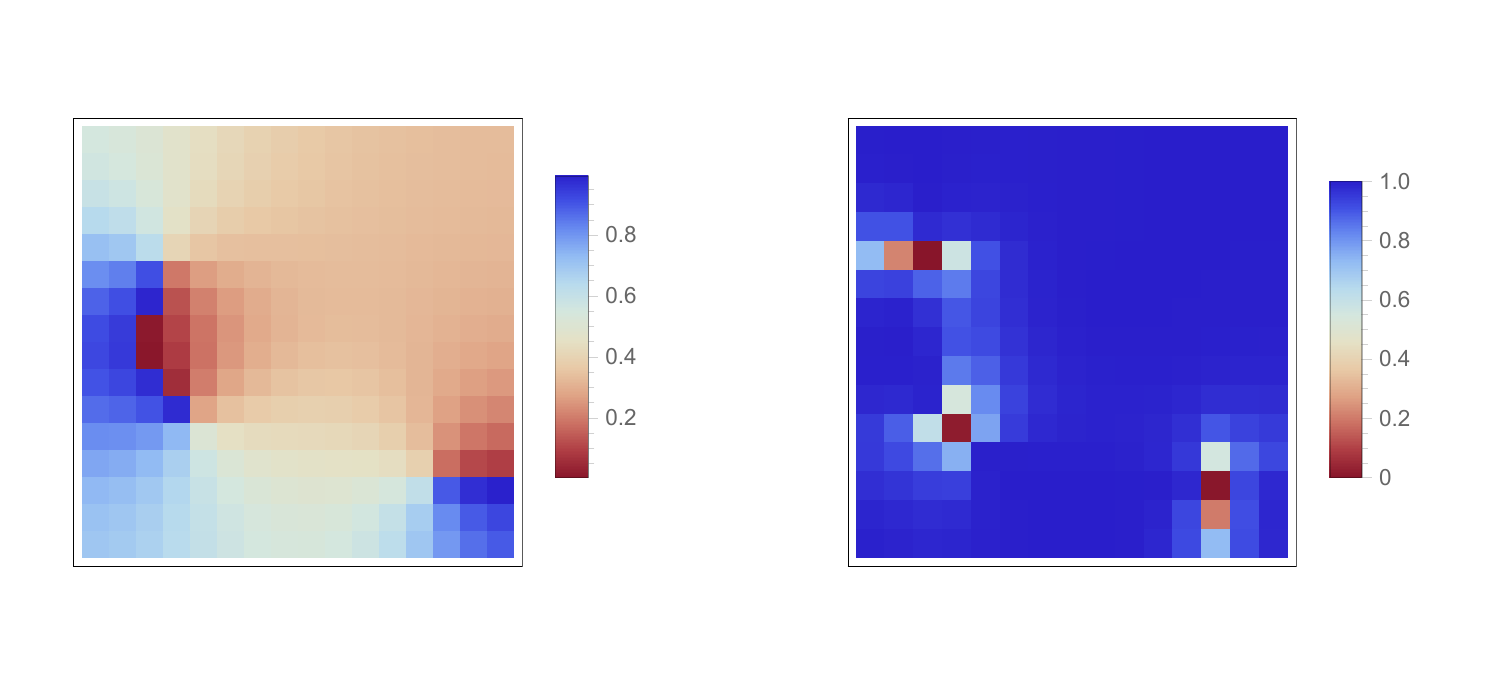}
  \caption{Multi-spiral waves in a $16\times 16$ grid. Left: Spiral waves in a $16\times 16$ grid. Right: Local order parameters. Red squares show the rotation center in Left. Simulation parameters: $n = 16, \omega_i = 1, k_{ij} = 1$.}
  \label{fig:resultIV7_2}
\end{figure}

Besides, multiple rotation centers can coexist, with adjacent rotation centers rotating in opposite directions (clockwise and anticlockwise), i.e., two rotation centers in a $8\times 8$ grid in FIG. \ref{fig:resultIV2_2 mul}C, and three rotation centers in a $16 \times 16$ grid in FIG. \ref{fig:resultIV7_2}. This property is also intuitively understandable: If there are two rotation centers A and B, and wave propagation direction between them is fixed, then the rotation direction relative to the two rotation centers is naturally opposite. 

\begin{figure}[htbp]
  \centering
  \includegraphics[width=0.8\textwidth]{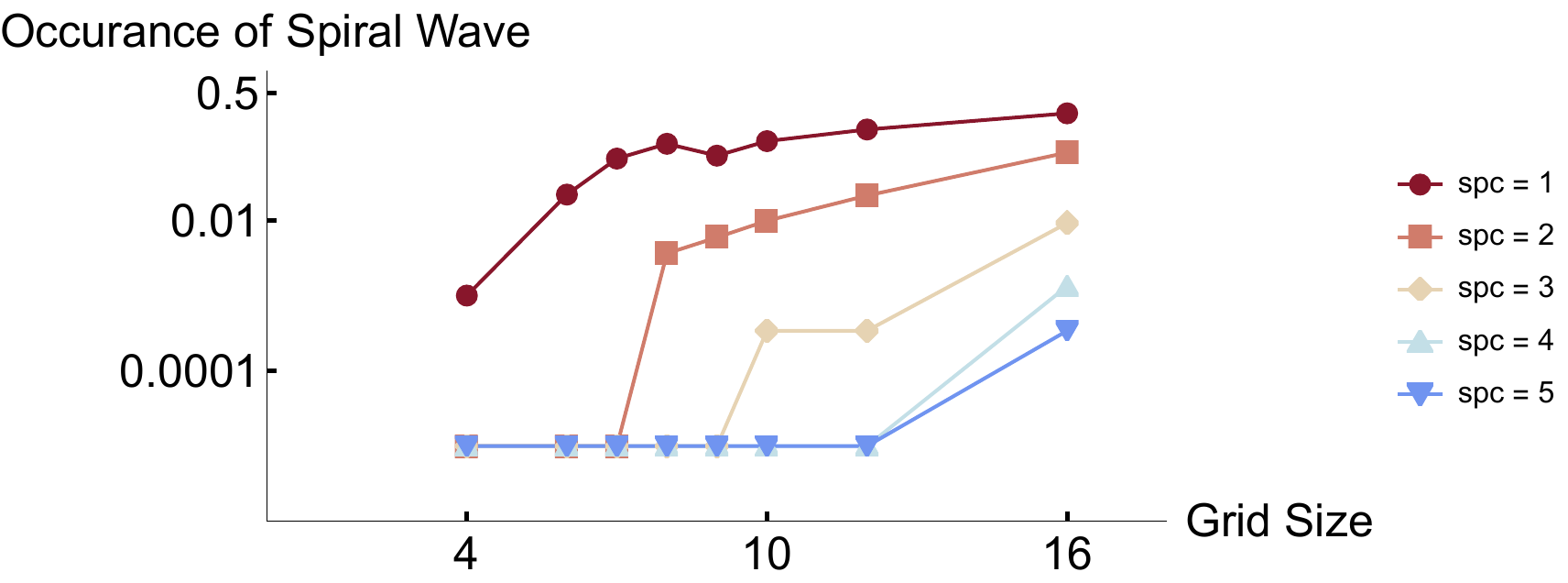}
  \caption{Occurrence of spiral wave changes with grid size. Different color: number of rotation centers. Simulation parameters: $n = 4,6,7,8,9,10,12,16, \omega_i = 1, k_{ij} = 1$.}
  \label{fig:resultIV6}
\end{figure}

Based on the numerical simulation results, the probability of spiral wave occurrence increases as the grid size increases. Additionally, multiple rotating centers are more likely to appear when the grid is relatively large.(FIG. \ref{fig:resultIV6}) It is consistent with \cite{Sarkar2021-sl}.

Another issue of interest is the robustness of the spiral wave pattern to noise. As discussed earlier, directional wave pattern is robust with limited noise, and similarly, spiral wave \textbf{W1} can also stably exist with noise (FIG. \ref{fig:increasing noise}B). Similarly, linearizing the system can obtain the first-order correction needed to maintain the original wave pattern at each phase, see eq.\ref{eq:error equilibrium condition}. However, there is a slight difference: at the rotation center, $ \cos(\theta^*_j-\theta^*_i) \approx 0$, so the equation satisfied by the first-order correction cannot be directly approximated by the original coupling strength. To quantitatively assess whether the first-order correction is a good approximation of the actual situation, an index $\frac{\Delta\theta_L - \Delta\theta_o}{\Delta\theta_o}$ is calculated for each oscillator in the grid. Here, $\Delta\theta_o$ refers to the difference in phase between the stable state with noise added and the stable state without noise, while $\Delta\theta_L$ is its first-order correction. Numerical results show that the distribution of the index is concentrated mainly between 0 and 0.1, indicating that the first-order correction is sufficient to describe the effect of noise on the system.

\begin{figure}[htbp]
  \centering
  \includegraphics[width=1\textwidth]{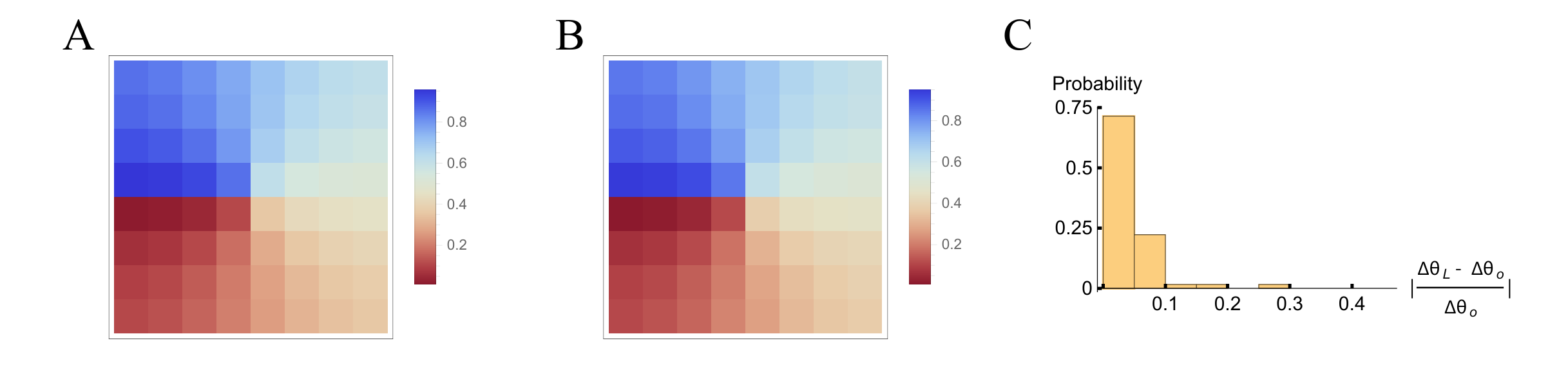}
  \caption{Small noise perturbation do not change the spiral wave pattern. A: Spiral wave in a $8\times 8$ grid. Simulation Paramter: $n = 8, \omega_i = 1, k_{ij} = 1$. B: Spiral wave in a $8 \times 8$ grid with small noise on $\omega_i$ and $k_{ij}$, in which $\omega_i \sim \mathcal{N}(1,0.05), k_{ij} \sim \mathcal{N}(1,0.05)$. Initial condition is the same to that used in A. C: Distribution of $\frac{\Delta\theta_L - \Delta\theta_o}{\Delta\theta_o}$ in B.}
  \label{fig:resultIV2_2}
\end{figure}

\section{Conclusion}\label{sec:conc}

This article investigates the properties of spatial waves formed by locally coupled oscillators in an $n \times n$ grid in the Kuramoto model. Numerical simulations show that the system will form directional waves in the presence of heterogeneity, while spiral waves may form in a system composed of homogeneous oscillators. Both waveforms remain stable under small noise interference. By analytically calculating the phase difference between oscillators in a $2 \times 2$ grid, we discuss the formation of waveforms in the system and then apply this method to larger grids. We analytically calculate the stable and saddle points of the system and discuss the stability of their corresponding waveforms. Linear approximation further demonstrates that waveforms with noise are the waveforms without noise plus first-order approximation, so the waveforms remain unchanged within a specific range of noise. These results suggest that in natural biological systems, the necessary condition for directional wave propagation is the presence of heterogeneity far beyond the noise, while the disappearance of heterogeneity may induce spiral waves, which often correspond to disease states. This article discusses a simple arrangement of oscillator systems while existing systems are often more complex. For example, the shape of delta cells in the pancreas determines that they can be electrically coupled to more cells, so connectivity in space may also be heterogeneous, and the spatial arrangement of cells is often spherical or ellipsoidal. Hence, the influence of three-dimensional spatial structures on wave pattern formation must be considered. On the other hand, the coupling function discussed in the Kuramoto model is the simplest trigonometric function $\sin(x)$. Investigating how the conclusion will change if changing the coupling function to other periodic functions is also essential.

\bibliography{ref}

\begin{thebibliography}{15}%
\makeatletter
\providecommand \@ifxundefined [1]{%
 \@ifx{#1\undefined}
}%
\providecommand \@ifnum [1]{%
 \ifnum #1\expandafter \@firstoftwo
 \else \expandafter \@secondoftwo
 \fi
}%
\providecommand \@ifx [1]{%
 \ifx #1\expandafter \@firstoftwo
 \else \expandafter \@secondoftwo
 \fi
}%
\providecommand \natexlab [1]{#1}%
\providecommand \enquote  [1]{``#1''}%
\providecommand \bibnamefont  [1]{#1}%
\providecommand \bibfnamefont [1]{#1}%
\providecommand \citenamefont [1]{#1}%
\providecommand \href@noop [0]{\@secondoftwo}%
\providecommand \href [0]{\begingroup \@sanitize@url \@href}%
\providecommand \@href[1]{\@@startlink{#1}\@@href}%
\providecommand \@@href[1]{\endgroup#1\@@endlink}%
\providecommand \@sanitize@url [0]{\catcode `\\12\catcode `\$12\catcode
  `\&12\catcode `\#12\catcode `\^12\catcode `\_12\catcode `\%12\relax}%
\providecommand \@@startlink[1]{}%
\providecommand \@@endlink[0]{}%
\providecommand \url  [0]{\begingroup\@sanitize@url \@url }%
\providecommand \@url [1]{\endgroup\@href {#1}{\urlprefix }}%
\providecommand \urlprefix  [0]{URL }%
\providecommand \Eprint [0]{\href }%
\providecommand \doibase [0]{https://doi.org/}%
\providecommand \selectlanguage [0]{\@gobble}%
\providecommand \bibinfo  [0]{\@secondoftwo}%
\providecommand \bibfield  [0]{\@secondoftwo}%
\providecommand \translation [1]{[#1]}%
\providecommand \BibitemOpen [0]{}%
\providecommand \bibitemStop [0]{}%
\providecommand \bibitemNoStop [0]{.\EOS\space}%
\providecommand \EOS [0]{\spacefactor3000\relax}%
\providecommand \BibitemShut  [1]{\csname bibitem#1\endcsname}%
\let\auto@bib@innerbib\@empty
\bibitem [{\citenamefont {Winfree}(1967)}]{Winfree1967-mr}%
  \BibitemOpen
  \bibfield  {author} {\bibinfo {author} {\bibfnamefont {A.~T.}\ \bibnamefont
  {Winfree}},\ }\bibfield  {title} {\bibinfo {title} {Biological rhythms and
  the behavior of populations of coupled oscillators},\ }\href@noop {}
  {\bibfield  {journal} {\bibinfo  {journal} {J Theor Biol}\ }\textbf {\bibinfo
  {volume} {16}},\ \bibinfo {pages} {15} (\bibinfo {year} {1967})}\BibitemShut
  {NoStop}%
\bibitem [{\citenamefont {Strogatz}(2000)}]{Strogatz2000-ci}%
  \BibitemOpen
  \bibfield  {author} {\bibinfo {author} {\bibfnamefont {S.~H.}\ \bibnamefont
  {Strogatz}},\ }\bibfield  {title} {\bibinfo {title} {From kuramoto to
  crawford: exploring the onset of synchronization in populations of coupled
  oscillators},\ }\href@noop {} {\bibfield  {journal} {\bibinfo  {journal}
  {Physica D: Nonlinear Phenomena}\ }\textbf {\bibinfo {volume} {143}},\
  \bibinfo {pages} {1} (\bibinfo {year} {2000})}\BibitemShut {NoStop}%
\bibitem [{\citenamefont {Acebr{\'o}n}\ \emph {et~al.}(2005)\citenamefont
  {Acebr{\'o}n}, \citenamefont {Bonilla}, \citenamefont {P{\'e}rez~Vicente},
  \citenamefont {Ritort},\ and\ \citenamefont {Spigler}}]{Acebron2005-wz}%
  \BibitemOpen
  \bibfield  {author} {\bibinfo {author} {\bibfnamefont {J.~A.}\ \bibnamefont
  {Acebr{\'o}n}}, \bibinfo {author} {\bibfnamefont {L.~L.}\ \bibnamefont
  {Bonilla}}, \bibinfo {author} {\bibfnamefont {C.~J.}\ \bibnamefont
  {P{\'e}rez~Vicente}}, \bibinfo {author} {\bibfnamefont {F.}~\bibnamefont
  {Ritort}},\ and\ \bibinfo {author} {\bibfnamefont {R.}~\bibnamefont
  {Spigler}},\ }\bibfield  {title} {\bibinfo {title} {The kuramoto model: A
  simple paradigm for synchronization phenomena},\ }\href@noop {} {\bibfield
  {journal} {\bibinfo  {journal} {Reviews of Modern Physics}\ }\textbf
  {\bibinfo {volume} {77}},\ \bibinfo {pages} {137} (\bibinfo {year}
  {2005})}\BibitemShut {NoStop}%
\bibitem [{\citenamefont {Ermentrout}\ and\ \citenamefont
  {Kleinfeld}(2001)}]{Ermentrout2001-ar}%
  \BibitemOpen
  \bibfield  {author} {\bibinfo {author} {\bibfnamefont {G.~B.}\ \bibnamefont
  {Ermentrout}}\ and\ \bibinfo {author} {\bibfnamefont {D.}~\bibnamefont
  {Kleinfeld}},\ }\bibfield  {title} {\bibinfo {title} {Traveling electrical
  waves in cortex: insights from phase dynamics and speculation on a
  computational role},\ }\href@noop {} {\bibfield  {journal} {\bibinfo
  {journal} {Neuron}\ }\textbf {\bibinfo {volume} {29}},\ \bibinfo {pages} {33}
  (\bibinfo {year} {2001})}\BibitemShut {NoStop}%
\bibitem [{\citenamefont {Bers}(2008)}]{Bers2008}%
  \BibitemOpen
  \bibfield  {author} {\bibinfo {author} {\bibfnamefont {D.~M.}\ \bibnamefont
  {Bers}},\ }\bibfield  {title} {\bibinfo {title} {Calcium cycling and
  signaling in cardiac myocytes},\ }\href
  {https://doi.org/10.1146/annurev.physiol.70.113006.100455} {\bibfield
  {journal} {\bibinfo  {journal} {Annual Review of Physiology}\ }\textbf
  {\bibinfo {volume} {70}},\ \bibinfo {pages} {23} (\bibinfo {year} {2008})},\
  \bibinfo {note} {pMID: 17988210},\ \Eprint
  {https://arxiv.org/abs/https://doi.org/10.1146/annurev.physiol.70.113006.100455}
  {https://doi.org/10.1146/annurev.physiol.70.113006.100455} \BibitemShut
  {NoStop}%
\bibitem [{\citenamefont {Ren}\ \emph {et~al.}(2022)\citenamefont {Ren},
  \citenamefont {Li}, \citenamefont {Han}, \citenamefont {Yu}, \citenamefont
  {Shi}, \citenamefont {Peng}, \citenamefont {Zhang}, \citenamefont {Wu},
  \citenamefont {Yang}, \citenamefont {Kim}, \citenamefont {Chen},\ and\
  \citenamefont {Tang}}]{Ren2022-bo}%
  \BibitemOpen
  \bibfield  {author} {\bibinfo {author} {\bibfnamefont {H.}~\bibnamefont
  {Ren}}, \bibinfo {author} {\bibfnamefont {Y.}~\bibnamefont {Li}}, \bibinfo
  {author} {\bibfnamefont {C.}~\bibnamefont {Han}}, \bibinfo {author}
  {\bibfnamefont {Y.}~\bibnamefont {Yu}}, \bibinfo {author} {\bibfnamefont
  {B.}~\bibnamefont {Shi}}, \bibinfo {author} {\bibfnamefont {X.}~\bibnamefont
  {Peng}}, \bibinfo {author} {\bibfnamefont {T.}~\bibnamefont {Zhang}},
  \bibinfo {author} {\bibfnamefont {S.}~\bibnamefont {Wu}}, \bibinfo {author}
  {\bibfnamefont {X.}~\bibnamefont {Yang}}, \bibinfo {author} {\bibfnamefont
  {S.}~\bibnamefont {Kim}}, \bibinfo {author} {\bibfnamefont {L.}~\bibnamefont
  {Chen}},\ and\ \bibinfo {author} {\bibfnamefont {C.}~\bibnamefont {Tang}},\
  }\bibfield  {title} {\bibinfo {title} {Pancreatic $\alpha$ and $\beta$ cells
  are globally phase-locked},\ }\href@noop {} {\bibfield  {journal} {\bibinfo
  {journal} {Nature Communications}\ }\textbf {\bibinfo {volume} {13}}
  (\bibinfo {year} {2022})}\BibitemShut {NoStop}%
\bibitem [{\citenamefont {Hoang}\ \emph {et~al.}(2015)\citenamefont {Hoang},
  \citenamefont {Jo},\ and\ \citenamefont {Hong}}]{Hoang2015-bh}%
  \BibitemOpen
  \bibfield  {author} {\bibinfo {author} {\bibfnamefont {D.-T.}\ \bibnamefont
  {Hoang}}, \bibinfo {author} {\bibfnamefont {J.}~\bibnamefont {Jo}},\ and\
  \bibinfo {author} {\bibfnamefont {H.}~\bibnamefont {Hong}},\ }\bibfield
  {title} {\bibinfo {title} {Traveling wave in a three-dimensional array of
  conformist and contrarian oscillators},\ }\href@noop {} {\bibfield  {journal}
  {\bibinfo  {journal} {Physical Review E}\ }\textbf {\bibinfo {volume} {91}}
  (\bibinfo {year} {2015})}\BibitemShut {NoStop}%
\bibitem [{\citenamefont {Bick}\ \emph {et~al.}(2020)\citenamefont {Bick},
  \citenamefont {Goodfellow}, \citenamefont {Laing},\ and\ \citenamefont
  {Martens}}]{Bick2020-mi}%
  \BibitemOpen
  \bibfield  {author} {\bibinfo {author} {\bibfnamefont {C.}~\bibnamefont
  {Bick}}, \bibinfo {author} {\bibfnamefont {M.}~\bibnamefont {Goodfellow}},
  \bibinfo {author} {\bibfnamefont {C.~R.}\ \bibnamefont {Laing}},\ and\
  \bibinfo {author} {\bibfnamefont {E.~A.}\ \bibnamefont {Martens}},\
  }\bibfield  {title} {\bibinfo {title} {Understanding the dynamics of
  biological and neural oscillator networks through exact mean-field
  reductions: a review},\ }\href@noop {} {\bibfield  {journal} {\bibinfo
  {journal} {The Journal of Mathematical Neuroscience}\ }\textbf {\bibinfo
  {volume} {10}} (\bibinfo {year} {2020})}\BibitemShut {NoStop}%
\bibitem [{\citenamefont {Paullet}\ and\ \citenamefont
  {Ermentrout}(1994)}]{Paullet1994}%
  \BibitemOpen
  \bibfield  {author} {\bibinfo {author} {\bibfnamefont {J.~E.}\ \bibnamefont
  {Paullet}}\ and\ \bibinfo {author} {\bibfnamefont {G.~B.}\ \bibnamefont
  {Ermentrout}},\ }\bibfield  {title} {\bibinfo {title} {Stable rotating waves
  in two-dimensional discrete active media},\ }\href
  {https://doi.org/10.1137/S0036139993250683} {\bibfield  {journal} {\bibinfo
  {journal} {SIAM Journal on Applied Mathematics}\ }\textbf {\bibinfo {volume}
  {54}},\ \bibinfo {pages} {1720} (\bibinfo {year} {1994})},\ \Eprint
  {https://arxiv.org/abs/https://doi.org/10.1137/S0036139993250683}
  {https://doi.org/10.1137/S0036139993250683} \BibitemShut {NoStop}%
\bibitem [{\citenamefont {{\'A}valos}\ \emph {et~al.}(2009)\citenamefont
  {{\'A}valos}, \citenamefont {Lai},\ and\ \citenamefont
  {Chan}}]{Avalos2009-nc}%
  \BibitemOpen
  \bibfield  {author} {\bibinfo {author} {\bibfnamefont {E.}~\bibnamefont
  {{\'A}valos}}, \bibinfo {author} {\bibfnamefont {P.-Y.}\ \bibnamefont
  {Lai}},\ and\ \bibinfo {author} {\bibfnamefont {C.~K.}\ \bibnamefont
  {Chan}},\ }\bibfield  {title} {\bibinfo {title} {Zero-refractoriness spirals
  in phase-coupled excitable media},\ }\href@noop {} {\bibfield  {journal}
  {\bibinfo  {journal} {Physical Review E}\ }\textbf {\bibinfo {volume} {80}}
  (\bibinfo {year} {2009})}\BibitemShut {NoStop}%
\bibitem [{\citenamefont {Sieber}\ and\ \citenamefont
  {Kalm{\'a}r-Nagy}(2011)}]{Sieber2011-mr}%
  \BibitemOpen
  \bibfield  {author} {\bibinfo {author} {\bibfnamefont {J.}~\bibnamefont
  {Sieber}}\ and\ \bibinfo {author} {\bibfnamefont {T.}~\bibnamefont
  {Kalm{\'a}r-Nagy}},\ }\bibfield  {title} {\bibinfo {title} {Stability of a
  chain of phase oscillators},\ }\href@noop {} {\bibfield  {journal} {\bibinfo
  {journal} {Physical Review E}\ }\textbf {\bibinfo {volume} {84}} (\bibinfo
  {year} {2011})}\BibitemShut {NoStop}%
\bibitem [{\citenamefont {Udeigwe}\ and\ \citenamefont
  {Ermentrout}(2015)}]{Udeigwe2015}%
  \BibitemOpen
  \bibfield  {author} {\bibinfo {author} {\bibfnamefont {L.~C.}\ \bibnamefont
  {Udeigwe}}\ and\ \bibinfo {author} {\bibfnamefont {G.~B.}\ \bibnamefont
  {Ermentrout}},\ }\bibfield  {title} {\bibinfo {title} {Waves and patterns on
  regular graphs},\ }\href {https://doi.org/10.1137/140969488} {\bibfield
  {journal} {\bibinfo  {journal} {SIAM Journal on Applied Dynamical Systems}\
  }\textbf {\bibinfo {volume} {14}},\ \bibinfo {pages} {1102} (\bibinfo {year}
  {2015})},\ \Eprint {https://arxiv.org/abs/https://doi.org/10.1137/140969488}
  {https://doi.org/10.1137/140969488} \BibitemShut {NoStop}%
\bibitem [{\citenamefont {Bramburger}(2019)}]{Jason2019}%
  \BibitemOpen
  \bibfield  {author} {\bibinfo {author} {\bibfnamefont {J.}~\bibnamefont
  {Bramburger}},\ }\bibfield  {title} {\bibinfo {title} {Stability of infinite
  systems of coupled oscillators via random walks on weighted graphs},\ }\href
  {https://doi.org/10.1090/tran/7609} {\bibfield  {journal} {\bibinfo
  {journal} {Transactions of the American Mathematical Society}\ }\textbf
  {\bibinfo {volume} {372}},\ \bibinfo {pages} {1159} (\bibinfo {year}
  {2019})}\BibitemShut {NoStop}%
\bibitem [{\citenamefont {Sarkar}\ and\ \citenamefont
  {Gupte}(2021)}]{Sarkar2021-sl}%
  \BibitemOpen
  \bibfield  {author} {\bibinfo {author} {\bibfnamefont {M.}~\bibnamefont
  {Sarkar}}\ and\ \bibinfo {author} {\bibfnamefont {N.}~\bibnamefont {Gupte}},\
  }\bibfield  {title} {\bibinfo {title} {Phase synchronization in the
  two-dimensional kuramoto model: Vortices and duality},\ }\href@noop {}
  {\bibfield  {journal} {\bibinfo  {journal} {Physical Review E}\ }\textbf
  {\bibinfo {volume} {103}} (\bibinfo {year} {2021})}\BibitemShut {NoStop}%
\bibitem [{\citenamefont {Tilles}\ \emph {et~al.}(2011)\citenamefont {Tilles},
  \citenamefont {Ferreira},\ and\ \citenamefont {Cerdeira}}]{Tilles2011-mp}%
  \BibitemOpen
  \bibfield  {author} {\bibinfo {author} {\bibfnamefont {P.~F.~C.}\
  \bibnamefont {Tilles}}, \bibinfo {author} {\bibfnamefont {F.~F.}\
  \bibnamefont {Ferreira}},\ and\ \bibinfo {author} {\bibfnamefont {H.~A.}\
  \bibnamefont {Cerdeira}},\ }\bibfield  {title} {\bibinfo {title} {Multistable
  behavior above synchronization in a locally coupled kuramoto model},\
  }\href@noop {} {\bibfield  {journal} {\bibinfo  {journal} {Physical Review
  E}\ }\textbf {\bibinfo {volume} {83}} (\bibinfo {year} {2011})}\BibitemShut
  {NoStop}%
\end{thebibliography}%

\section{Appendix}\label{sec:Appendix}

\subsection{Proof of Lemma 1}

\textbf{Lemma 1} Dynamic system (eq.1) reaches an equilibrium state
equivalent to that all oscillators \(\theta_i\) are phase-locked to
average rotation \(\overline{\omega}t\). In other words,
\(\theta_i -\overline{\omega}t \equiv \Omega_i, \forall i\).

\textbf{Proof}: It is evident that the condition is sufficient since the
phase difference is a constant:

\begin{equation}
 \begin{aligned}
 \theta_j - \theta_i &= (\theta_j-\overline{\omega} t)- (\theta_i-\overline{\omega} t)\\ 
 &= \Omega_j - \Omega_i
 \end{aligned} 
\end{equation}

Now to prove its necessity.

Since \(\theta_j -\theta_i \equiv C_{ji}, \forall i,j \), then it is
equivalent to:

\be
 \frac{d(\theta_j - \theta_i)}{dt} = 0 \Leftrightarrow \frac{d\theta_j}{dt} = \frac{d\theta_i}{dt} , \forall i,j
\ee

Thus, all oscillators have the same velocity, denoted as \(v(t)\).

On the other hand, the average speed of all oscillators is equal to
\(\overline{\omega_i}\) due to

\be
 \sum_i\sum_{j\in \Lambda} K_{ji}\sin(\theta_j-\theta_i) = 0
\ee

, which is a result of symmetrical coupling.

Therefore \(v(t) = \overline{\omega}\) and it indicates

\be
 \frac{d(\theta_i-\overline{\omega}t)}{dt} = 0
\ee

It concludes that all oscillators \(\theta_i\) are phase-locked to
average rotation \(\overline{\omega}t\). 

\subsection{Explicit form of the null vector}

Here is the explicit form of $c_1,c_4$ in eq.\ref{eq:cos}:
\begin{align}
 c_1 &= 3 w_1^3+w_2^3+w_3^3+w_3^2 w_4+3 w_3 w_4^2 \notag \\
 &+3 w_4^3+ 8 k w_3^2 \delta+8 k w_4^2 \delta \notag \\ 
 &-w_1^2(13 w_2+w_3-5 w_4+ 8 k \delta) \notag \\
 &-w_2^2(w_3+3 w_4+8 k \delta)\\
 &-w_2(w_3^2+6 w_3 w_4+w_4^2+16 k w_3 \delta) \notag \\ 
 &+w_1(w_2^2-3 w_3^2-2 w_3 w_4-11 w_4^2 \notag \\
 &-16 k w_4 \delta+2 w_2(5 w_3+7 w_4+16 k \delta)) \notag \\
 c_4 &= 3 w_1^3-w_2^3+w_3^3-w_3^2 w_4-5 w_3 w_4^2 \notag \\
 &-3 w_4^3+w_2(w_3+w_4)^2 \notag \\
 &+8 k w_3^2 \delta-16 k w_3 w_4 \delta-8 k w_4^2 \delta \notag \\
 &+w_1^2(5 w_2-w_3-13 w_4-8 k \delta)\\
 &+w_2^2(-w_3+3 w_4+8 k \delta) \notag \\
 &+w_1(w_2^2-3 w_3^2+10 w_3 w_4+13 w_4^2 \notag \\
 &+32 k w_4 \delta-2 w_2(w_3+5 w_4+8 k \delta)) \notag
\end{align}

\end{document}